\documentclass[%
 reprint,
superscriptaddress,
showpacs,
amsmath,amssymb,aps,prb,]{revtex4-1}

\usepackage{microtype}
\usepackage{graphicx}
\usepackage{dcolumn}
\usepackage{bm}
\usepackage{hyperref}
\usepackage{braket}

\newcommand{\ed}[1]{e^\dagger_{#1}}
\newcommand{\e}[1]{e_{#1}}
\newcommand{\hd}[1]{h^\dagger_{\bar{#1}}}
\newcommand{\h}[1]{h_{\bar{#1}}}

\newcommand{\ad}{a^ \dagger}

\begin{document}


\title{Magnetic control of dipolaritons in quantum dots}
\author{J. S. \surname{Rojas-Arias}}
\email{jusrojasar@unal.edu.co}
\affiliation{Departamento de F\'isica, Universidad Nacional de Colombia, Bogot\'a D.C., Colombia}
\author{B. A. \surname{Rodr\'iguez}}
\affiliation{
Instituto de F\'isica, Universidad de Antioquia, Medell\'in, Colombia}
\author{H. \surname{Vinck-Posada}}
\affiliation{Departamento de F\'isica, Universidad Nacional de Colombia, Bogot\'a D.C., Colombia}

\date{\today}

\begin{abstract}

Dipolaritons are quasiparticles that arise in coupled quantum wells embedded in a microcavity, they are a superposition of a photon, a direct exciton and an indirect exciton. We propose the existence of dipolaritons in a system of two coupled quantum dots inside a microcavity in direct analogy with the quantum well case and find that, despite some similarities, dipolaritons in quantum dots have different properties and can lead to true dark polariton states. We use a finite system theory to study the effects of the magnetic field on the system, including the emission, and find that it can be used as a control parameter of the properties of excitons and dipolaritons, and the overall magnetic behaviour of the structure.
\end{abstract}

\maketitle

\section{\label{sec:introduction}Introduction}

Microcavities with active media embedded in them are structures of great interest in the development of quantum computation and the study of fundamental physics,\cite{Kavokin} these devices are the solid-state implementation of the Purcell effect.\cite{Purcell1946,Kleppner1981} Typical examples are quantum wells (QWs) or quantum dots (QDs) in micropillars or photonic crystals.\cite{Reithmaier2004,Robin2005,Reitzenstein2007,Englund2009} Exciton polaritons are quasiparticles that arise in this kind of systems when a strong coupling between matter excitations and cavity photons is obtained.\cite{Sanvitto,Laussy2009}  From their excitonic component, polaritons are provided with strong non-linear interactions, and a very small effective mass is inherited from their photonic contribution;\cite{Byrnes2014-2} these  and other characteristics make polaritons quasiparticles of high interest in research, leading to the development of the field known as \textit{polaritonics}.\cite{Koehl2001,Takahara2005,Feurer2007,Singh2009,DeLiberato2015}

The usage of double QWs as active media inside a microcavity leads to the formation of a new kind of polaritons.\cite{Christmann2010,Christmann2011,Cristofolini2012,Szymanska2012}. The QWs are electrically tuned to allow the tunnelling of charge carriers and the subsequent formation of bound electron-hole pairs with the charges confined in different layers, known as \textit{indirect excitons} (IX). The latter do not interact with cavity photons (C) but are electrically coupled with \textit{direct excitons} (DX), electron-hole pairs in the same QW. The three-way superposition of DX, IX and C is a \textit{dipolariton}, a polariton with a large dipole moment due to its indirect exciton part. Recently, dipolaritons in QW systems have been predicted to be suitable for the observation of superradiant terahertz (THz) emission and THz lasing,\cite{Kristinsson2013,Kyriienko2013,Kristinsson2014} single-photon emission, \cite{Kyriienko2014,Kyriienko2014-2} Bose-Einstein condensation,\cite{Byrnes2014,Su2014} polariton bistability,\cite{Coulson2013} and the realization of high-fidelity quantum gates.\cite{Kyriienko2016} Meanwhile, other research has focused on the effects of the pumping,\cite{Khadzhi2015} and the electrical control of strong coupling\cite{Sivalertporn2015} and spin interactions\cite{Nalitov2014}.

\begin{figure}[tb]
\centering
\includegraphics[width=5.5cm]{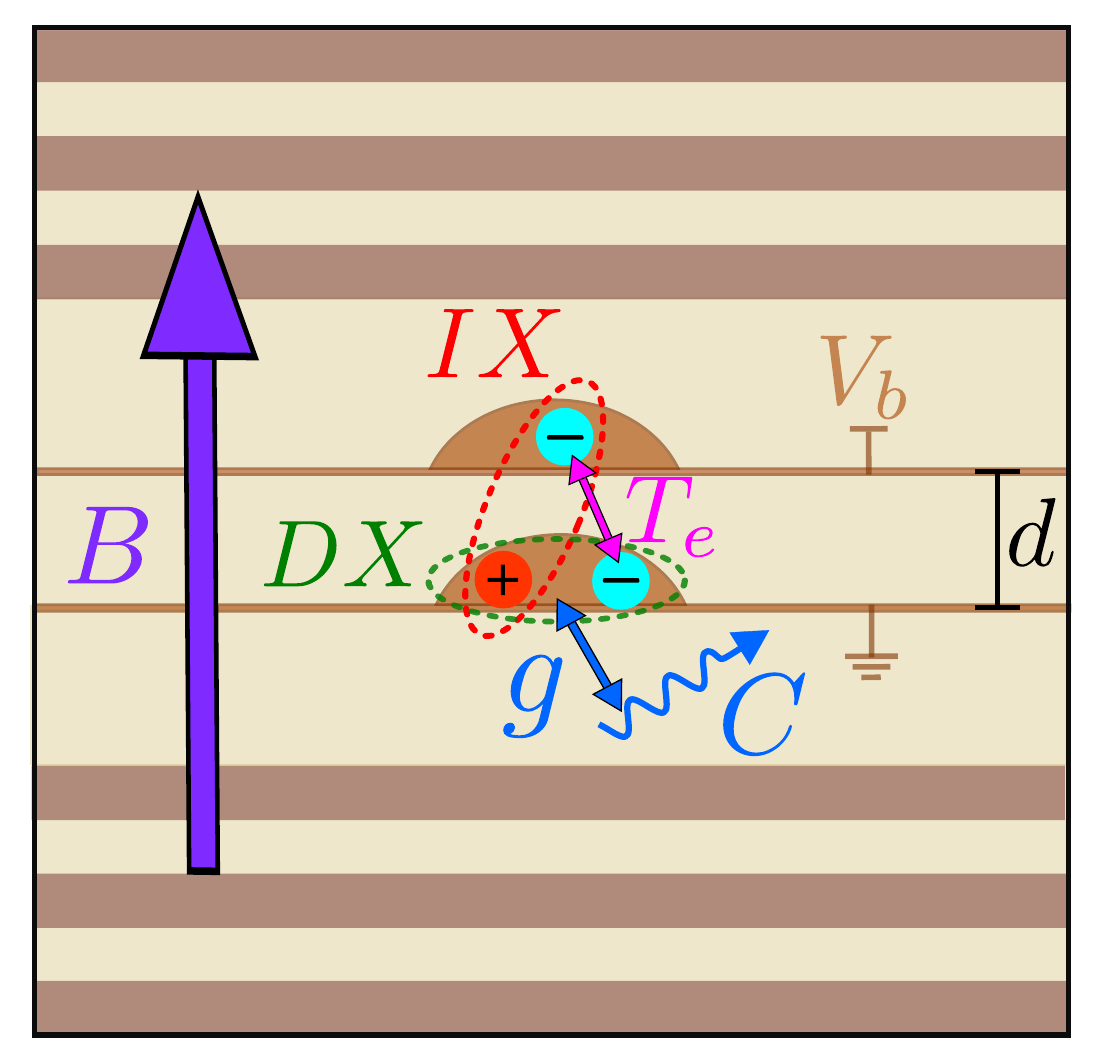}
\caption{(Color Online) Scheme of a QDM, composed of two coupled QDs separated by a distance $d$, embedded in a microcavity. The tunnelling rate $T_e$ and bias voltage $V_b$ allows the formation of IXs.}
\label{fig:system}
\end{figure}

An analogue system can be fabricated replacing the QWs with QDs. As quantum dots are usually known as \textit{artificial atoms}, the coupling of 2 or more of them is called a \textit{quantum dot molecule} (QDM).\cite{Oosterkamp1998,Austing1998,Wu} Similarly to the QW setup, an electric field can be used to enhance the tunnelling coupling between dots and allow the formation of IXs.\cite{Villas-Boas2004,Villas-Boas2004-2,Muller2012,Muller2013} Thus, in principle, it is possible to generate dipolaritons in a QDM embedded in a microcavity (Fig. \ref{fig:system}). In this paper we explore this idea and study the differences and similarities between QD dipolaritons and their QW counterpart. Moreover, we include a constant magnetic field and propose it as a control channel of the properties of the system. Our main findings include the existence of true dark polariton states (despite the controversy around this states in QWs),\cite{Cristofolini2012,Szymanska2012,Sivalertporn2015} a scaling law for the energy of the system and a change in the magnetic response.

The paper is structured as follows. In Sec. \ref{sec:hamiltonian} we introduce the finite system hamiltonian used to describe the system of interest. The study of the properties of excitons in a QDM is presented in Sec. \ref{sec:exciton}. In Sec. \ref{sec:polariton} we propose the existence of dipolaritons in QDs and investigate their properties under the presence of a magnetic field. Finally, in Sec \ref{sec:multiexcitonic}, we present the extension to multiexcitonic QDs using a variational method. Sec. \ref{sec:conclusions} includes a summary and conclusions of the work.

\section{System and Hamiltonian \label{sec:hamiltonian}}

The system consists of a microcavity that confines two layers of vertically aligned InGaAs quantum dots, separated by a distance $d$, in the presence of a constant magnetic field $B$ in the growth direction of the structure,\cite{Rojas-Arias2015} as shown in Fig. \ref{fig:system}. The cavity is such that the energy levels are spaced enough allowing us to consider a single light mode. Only one QD is in strong coupling with the light field and the interaction is described upon dipolar and rotating wave approximations. For simplicity, we consider tunnelling between dots to be only of electrons, this is usually done experimentally by including a bias voltage $V_b$ that tunes the conduction band levels and detunes the valence bands.\cite{Muller2012,Villas-Boas2004} With the above considerations, one dot confines both electrons and holes (labelled as 1) and the other one confines electrons only (labelled as 2), such that the hamiltonian can be written as:

\begin{widetext}
\begin{equation}
\begin{aligned}
H&=\sum_{kn}(t_{kn}^{(e_1)}e_{1k}^{\dagger}e_{1n}+t_{\bar{k}\bar{n}}^{(h)}h_{\bar{k}}^{\dagger}h_{\bar{n}}+ t_{kn}^{(e_2)}e_{2k}^{\dagger}e_{2n})+\hbar\omega a^\dagger a+g\sum_n (a^\dagger h_{\bar{n}}e_{1n}+ae_{1n}^\dagger h_{\bar{n}}^\dagger )+T_e\sum_{n}(e_{1n}^\dagger e_{2n}+e_{2n}^\dagger e_{1n})\\
&+\dfrac{\beta}{2}\sum_{rsuv}\braket{rs|V(0)|uv}e_{1r}^\dagger e_{1s}^\dagger e_{1v}e_{1u} +\dfrac{\beta}{2}\sum_{rsuv}\braket{\bar{r}\bar{s}|V(0)|\bar{u}\bar{v}}h_{\bar{r}}^\dagger h_{\bar{s}}^\dagger h_{\bar{v}}h_{\bar{u}} +\dfrac{\beta}{2}\sum_{rsuv}\braket{rs|V(0)|uv}e_{2r}^\dagger e_{2s}^\dagger e_{2v}e_{2u}\\
&-\beta\sum_{rsuv}\braket{r\bar{s}|V(0)|u\bar{v}}e_{1r}^\dagger h_{\bar{s}}^\dagger h_{\bar{v}}e_{1u}+\beta\sum_{rsuv}\braket{rs|V(d)|uv}e_{1r}^\dagger e_{2s}^\dagger e_{2v}e_{1u}-\beta\sum_{rsuv}\braket{r\bar{s}|V(d)|u\bar{v}}e_{2r}^\dagger h_{\bar{s}}^\dagger h_{\bar{v}}e_{2u}.
\end{aligned}
\label{eq:Hamiltonian}
\end{equation}
\end{widetext}

The state $\ket{i}$ corresponds to a Landau level of an electron in a magnetic field, determined by radial $n_i$ and angular momentum $l_i$ quantum numbers. $\ket{\bar{i}}$ for holes represents the same state $\ket{i}$ for electrons except in the opposite sign of the angular momentum.\cite{Vinck2006} The single particle energies are ($\alpha=e_1,e_2,h$):

\begin{equation}
\begin{aligned}
t^{(\alpha)}_{ij}&=\dfrac{\hbar\omega_c^{\alpha}}{2}(2n_i\pm l_i+|l_i|+1)\delta_{i,j}\\
&+\dfrac{\hbar\omega_\alpha^2}{\omega_c^{\alpha}}\braket{i|r^2|j}+E_{gap}\delta_{\alpha,e}\delta_{i,j}+E_0^\alpha\delta_{i,j},
\label{tridiagonal}
\end{aligned}
\end{equation}

\noindent where $\omega_c^{\alpha}=eB/m_{\alpha}c$ is the cyclotron frequency, $m_e=0.05m_0$ and $m_h=0.07m_0$ are the effective masses with $m_0$ the free electron mass, $\omega_{\alpha}$ is the strength of the in-plane harmonic confinement inside the dots, $E_{gap}=850$meV is the energy gap of the material which is included in the electron energies only, and $E_0^\alpha$ is an offset that arises from the confinement in the $z$ direction and bias voltage.\cite{Riel2008,Villas-Boas2004} $\beta=2.94\sqrt{B}$meV is the strength of Coulomb interactions, with $B$ in Teslas.\cite{Vinck2006} $g$ is the Rabi coupling between direct excitons and the light field of energy $\hbar\omega$. $\braket{rs|V(d)|uv}$ is the matrix element of the Coulomb interaction between particles vertically separated by a distance $d$, the situation of charge carriers in the same QD corresponds to $d=0$.\cite{Mondragon-Shem2010} The tunnelling coupling is considered elastic\cite{Smoliner1991,Muller2012,Tarucha1999} with strength $T_e$. All the distances are in units of the magnetic length $l_B=\sqrt{2\hbar c/eB}$.

Spin degrees of freedom are ignored by considering a single polarization of light coupling with a specific spin orientation of electrons and holes, and by neglecting exchange effects. This argument is valid because tunnelling is considered elastic and Coulomb interaction does not change spin orientation.

\section{Excitons in a QDM \label{sec:exciton}}

To study the properties of excitons in QDMs we focus on the system without light i.e., $\hbar\omega=g=0$ in Eq. \eqref{eq:Hamiltonian}. 

InGaAs is a material with a direct band gap and, since the tunnelling is elastic, the total angular momentum is a conserved quantity.\cite{Vera2009} The number of excitons defined as:\cite{Eastham2001,Vinck2006,Vinck2007}

\begin{equation}
N_{exc}=\dfrac{1}{2}\sum_n(\ed{1n}\e{1n}+\ed{2n}\e{2n}+\hd{n}\h{n}),
\end{equation}

\noindent is another conserved quantity. Therefore, the hamiltonian is diagonalized for a fixed manifold, using a basis composed of states constructed as products: $\ket{S_{e1}}\ket{S_{e2}}\ket{S_h}$, with zero total angular momentum and equal number of electrons and holes.\cite{Vera2009,Vinck2006,Vinck2007} $\ket{S}$ represents a Slater determinant of single particle states of electrons in each dot or holes. The lowest eigenvalue will represent the ground-state energy and the corresponding linear combination of states used for the diagonalization will be the ground-state ket, similarly, the excited states are obtained.

It is important to notice that the Landau basis is infinite and must be truncated at a certain $l_{max}$ and $n_{max}$. This truncation is, in principle, due to computational reasons, however, it is also natural for the system since states with a large angular momentum and energy will overcome the confinement in the QD and leave. 

\begin{figure}[tb]
\centering
\includegraphics[width=9cm]{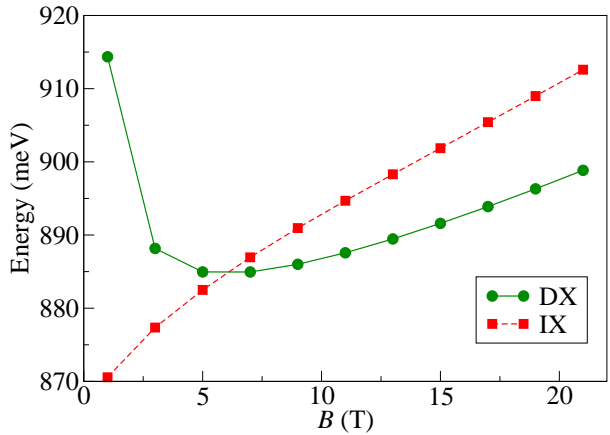}
\caption{(Color Online) Energy of the DX and IX as a function of the magnetic field.  The strengths of the parabolic confinements were taken as $\hbar\omega_h=8$meV, $\hbar\omega_{e_1}=12$meV and $\hbar\omega_{e_2}=18$meV, and the separation between dots was fixed at $10$nm. The corresponding energy offsets due to the confinement in the $z$ direction were set as $E_0^{h}=10$meV and $E_0^{e_1}=15$meV, and $E_0^{e_2}$ was set such that, by means of the bias voltage, the single particle electron states with quantum numbers $n=0$ and $l=0$, in both quantum dots, are tuned at the same energy.}
\label{fig:excitons energies}
\end{figure}

The diagonalization of the hamiltonian was performed using a basis constructed as explained above and composed of 120 single particle (Landau) states, distributed in 3 levels, i.e., $n_{max}=2$. We focused on a single exciton $N_{exc}=1$, however, the extension to several electron-hole pairs is straightforward. First, we ignore the tunnelling and plot the energies of the DX and IX for several values of magnetic field, the results are depicted in Fig. \ref{fig:excitons energies}. We see that while the IX energy increases monotonically with $B$, the DX energy has a richer behaviour. At low values of the magnetic field ($B<5T$) the Coulomb interaction gives a relevant contribution $\sim-\sqrt{B}$ that decreases the energy.\cite{Vinck2008} As the field increases, the contribution of Landau energies $\sim\hbar\omega_c\propto B$ becomes more relevant than Coulomb interactions and the energy starts increasing, causing a change in the slope. The IX has a simpler behaviour due to the spatial separation of the charge carriers that reduces the Coulomb matrix elements causing the effect of Coulomb's interaction to be almost lost.

\begin{figure}[tb]
\centering
\includegraphics[width=9.5cm]{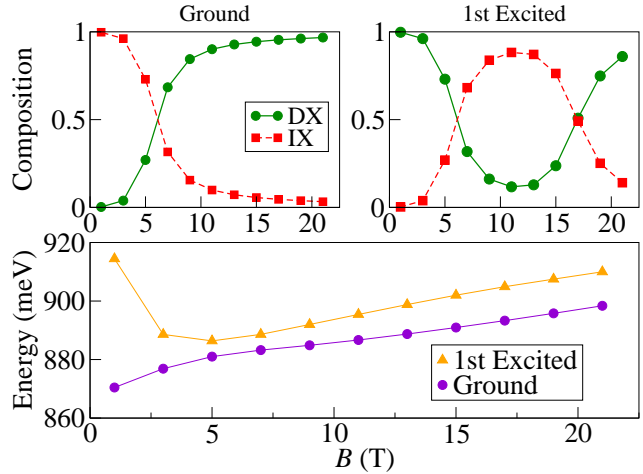}
\caption{(Color Online) Fractional composition and energy of the ground and first excited state, as a function of the magnetic field, for the system without light. The tunnelling rate was fixed at $T_e=3$meV, other parameters are the same as in Fig. \ref{fig:excitons energies}.}
\label{fig:system without light}
\end{figure}

In Fig. \ref{fig:system without light} we plot the energy and composition of the ground and first excited state of the tunnel-coupled system, for several values of the magnetic field. The expected avoided crossing between the energies of Fig. \ref{fig:excitons energies} is found. The composition of the ground state follows a behaviour that can be expected from the results in Fig. \ref{fig:excitons energies}. At small values of the magnetic field, for the used parameters, the IX has lower energy than the DX, for this reason the ground state is highly composed of IX. As the field increases, the DX and IX have similar energies and the ground state becomes a superposition of the two kinds of excitons. Finally, at high magnetic fields, the DX has lower energy compared to the IX and becomes the most relevant contribution on the composition. A similar analysis can be performed for the first excited state but there is something to note for $B>15$T. The IX composition of the state decreases while the DX counterpart increases, this occurs because the IX exceeds the energy of a higher excited state of the DX branch, DX(2s). The magnetic field provides an extra control channel that, jointly with the bias voltage, may be used to manipulate the IX fraction in the system, i.e., the overall dipole moment.

\section{Polaritons in QDMs \label{sec:polariton}}

In the previous section, the numerical diagonalization method was used to investigate both direct and indirect excitons' features. Now we extend this procedure to take light-matter interaction into account for the generation of polaritons with a large dipole moment.

We define dipolaritons in QDs as the dressed states of the hamiltonian \eqref{eq:Hamiltonian}. As in the QW case, they are a superposition of a DX, an IX and a cavity photon C; the IX provides the large dipole moment and a longer lifetime. The analogy with the QW case is straightforward, however, provided its fully quantisation and small number of excitations, one might expect quantum properties to be at its maximum expression in QDs; this is a fundamental discussion that still arises for usual polaritons\cite{Suarez2012} and we leave it for future research.

\begin{figure*}
\centering
\includegraphics[width=9cm]{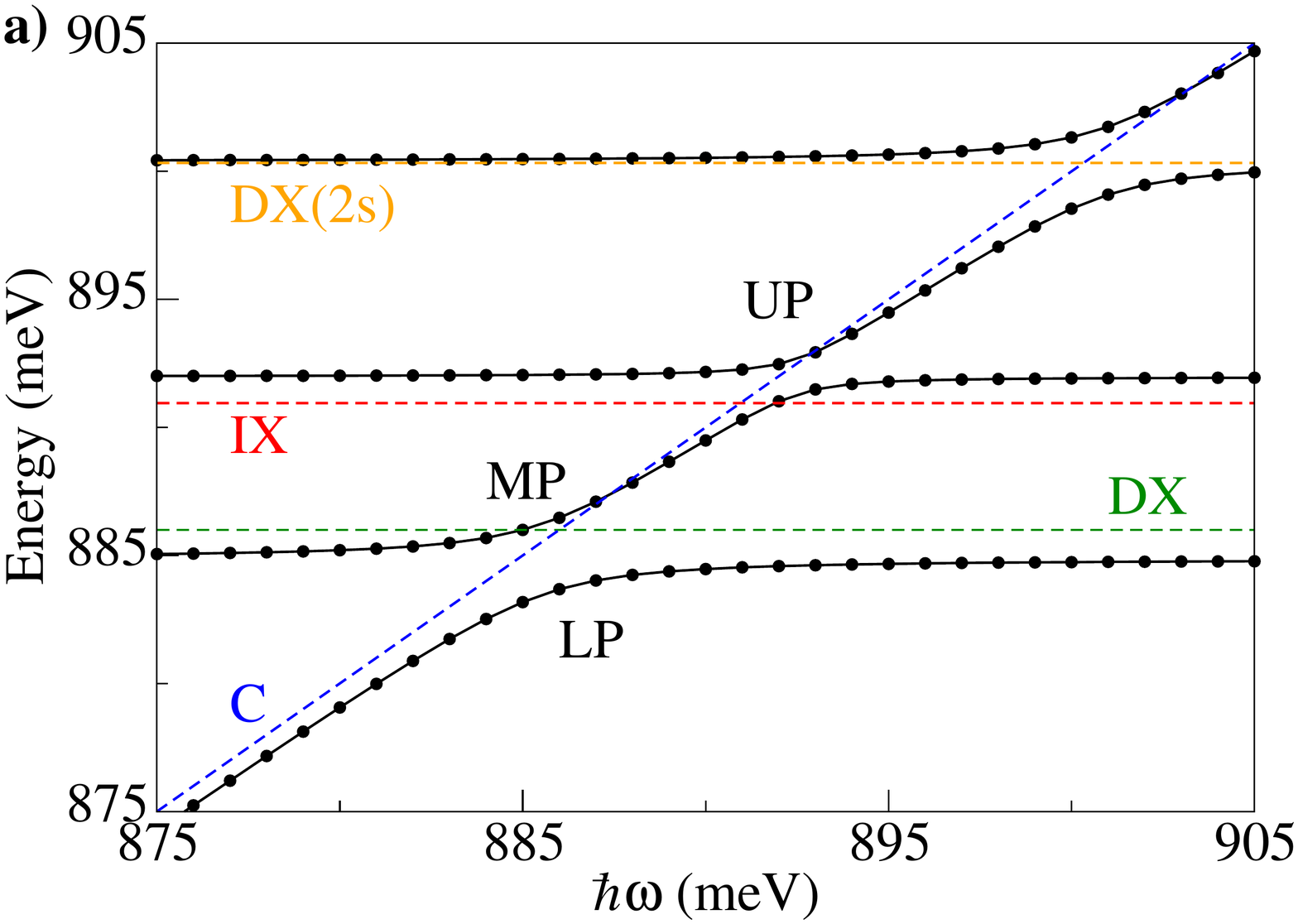}\includegraphics[width=9cm]{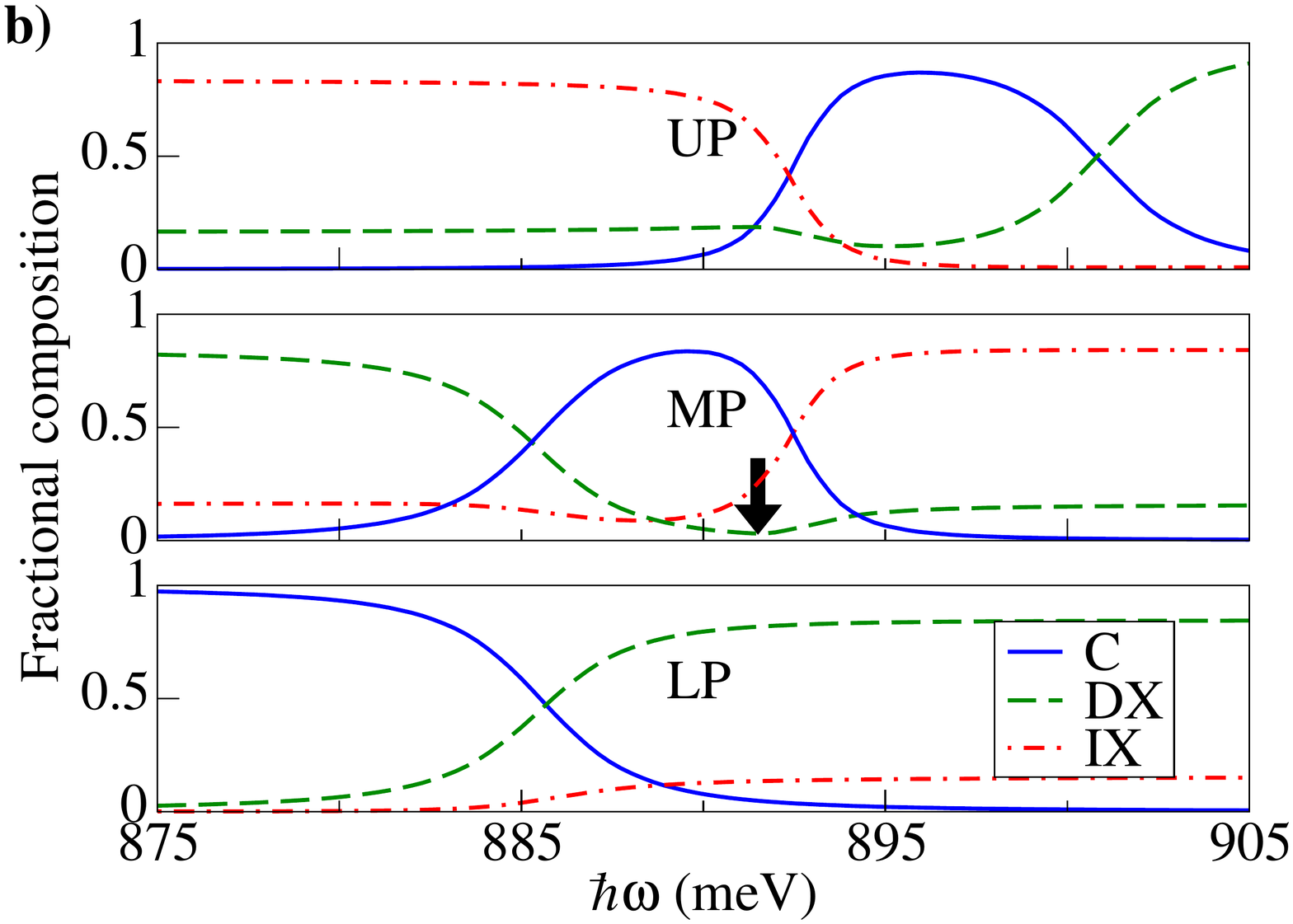}
\caption{(Color Online) a) First four eigenenergies of the hamiltonian from Eq. \eqref{eq:Hamiltonian} (black), bare energies of the photon (blue), indirect exciton (red), direct exciton (green) and a higher excited of the direct exciton (orange);  as a function of the photon energy. b) Composition of the three dipolariton branches as a function of the photon energy; black arrow in MP denotes the position of the dark polariton. $B=9$T and $g=1$meV, other parameters used are the same as in Fig. \ref{fig:system without light}.}
\label{fig:anticrossing}
\end{figure*}

When light is included, the hamiltonian commutes with the number of dipolaritons operator defined as:

\begin{equation}
N_{dip}=N_{ph}+N_{exc}=\ad a+N_{exc},
\end{equation}

\noindent again, the hamiltonian is diagonalized for a fixed manifold. We construct a basis similar to the one used in Sec. \ref{sec:exciton} but including a Fock state for the light: $\ket{N_{ph}}\ket{S_{e1}}\ket{S_{e2}}\ket{S_h}$. We study the situation with the lowest non-trivial manifold: $N_{dip}=1$.

\subsection{Dipolariton modes}

The first four eigenenergies of the system and the composition of the corresponding eigenstates as a function of the photon energy are plotted in Fig. \ref{fig:anticrossing}. The bare energies of the photon, indirect exciton and direct exciton are plotted with dashed lines as a reference. As in Ref. \onlinecite{Cristofolini2012}, the three dipolariton branches are recognized. The avoided crossing between the \textit{lower dipolariton} LP and \textit{middle dipolariton} MP branches occurs due to light-matter interaction. The anticrossing between the MP and the \textit{upper dipolariton} UP branches has the same reason, however, this splitting is reduced due to the tunnelling given the fact that IXs do not couple with light. From the composition of the dipolariton branches we can see that the optical detuning determines if the electron spends more time shuttling between QDs or Rabi flopping on the DX transition,\cite{Cristofolini2012} e.g., for $\hbar\omega>887$meV, the LP spends more time oscillating between DX and IX but for $\hbar\omega<887$meV, the LP tends to  perform Rabi oscillations.

Effective models like the ones used in Refs. \onlinecite{Cristofolini2012,Coulson2013,Kyriienko2013,Kristinsson2013, Kristinsson2014,Byrnes2014} cannot predict the avoided crossing between the UP and a higher energy branch shown in Fig. \ref{fig:anticrossing}a. Our finite system model allows us to have access to higher excited states of the excitons, in particular, of the direct exciton DX(2s), this anticrossing causes the UP to increase its composition of DX for high values of the photon energy (Fig. \ref{fig:anticrossing}b), exhibiting the limitations of effective models.

When the cavity photon and the indirect exciton have the same energy, the DX fraction is zero at the MP branch, therefore, the dipolariton becomes a mixture of C and IX that are not Rabi-coupled, this state is known as \textit{dark polariton} and is a prediction of the model used in Ref. \onlinecite{Cristofolini2012}. It was pointed out in Ref. \onlinecite{Sivalertporn2015} that this dark polariton state was just an artefact of the effective model that neglects higher excited states of the excitons, furthermore, the authors claim that no dark polariton is possible with QW dipolaritons. Our model including both the internal structure of excitons and their higher excited states, still predicts the dark polariton in the MP branch, showing that despite the similarity in the dispersion relations of QW and QD dipolaritons (Fig. \ref{fig:anticrossing}a), differences in system parameters lead to properties that distinguish both classes of polaritons, making worthwhile the study of the quantum dot situation.

\subsection{Effects of magnetic field on a single dipolariton}

\begin{figure}
\centering
\includegraphics[width=9.5cm]{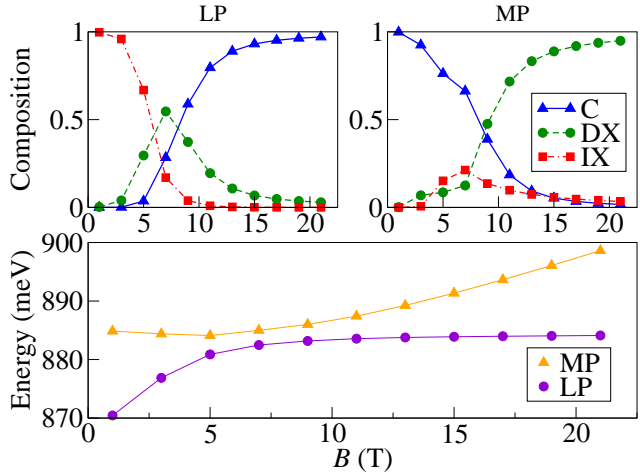}
\caption{(Color Online) Fractional composition and energy of the LP and MP branches, as a function of the magnetic field. The light energy was set to $\hbar\omega=885$meV.}
\label{fig:system with light}
\end{figure}

To see the effects of the magnetic field on a single dipolariton, we plot the energy of the ground state and the first excited state as a function of the magnetic field, the results are shown in Fig. \ref{fig:system with light}. Photon energy does not depend on the magnetic field while the energy of the superposition of DX and IX increases with the field; at $B\sim7$T both energies are near resonance and the light-matter interaction causes an avoided crossing.

The composition of the LP and the MP branches are plotted in the upper panels of Fig. \ref{fig:system with light}. At low $B$, the LP is completely composed of IX, in contrast, the MP is completely composed of light. Increasing the field, the composition of the states can be controlled to become a mixture of the three modes: IX, DX and C. Finally, at high values of the magnetic field, the matter acquires higher energy than light and the LP acquires a high photonic fraction.

This tuning of the composition of the dipolaritons with the magnetic field can be complemented with the bias voltage and the light mode energy (with a wedged cavity), thereby, becoming a highly controllable system.

\subsection{Emission}

Regardless of our non-dissipative model, the intensity of the emission can be obtained from the probability of annihilating a photon:\cite{Vinck2006}

\begin{align}
I\sim|\braket{N_{dip}-1|a|N_{dip}}|^2,
\label{eq:emission}
\end{align}

\noindent where $\ket{N_{dip}}$ and $\ket{N_{dip}-1}$ denote a state with dipolariton number $N_{dip}$ and $N_{dip}-1$, respectively. We assume that the system is operating at low temperatures, then the initial state is the ground state of the $N_{dip}$ system. From the computation of Eq. \eqref{eq:emission} we find that the transition to the ground state of the $N_{dip}-1$ system is the only one with a significant contribution to the intensity.\cite{Vinck2006}

\begin{figure}
\centering
\includegraphics[width=6.5cm]{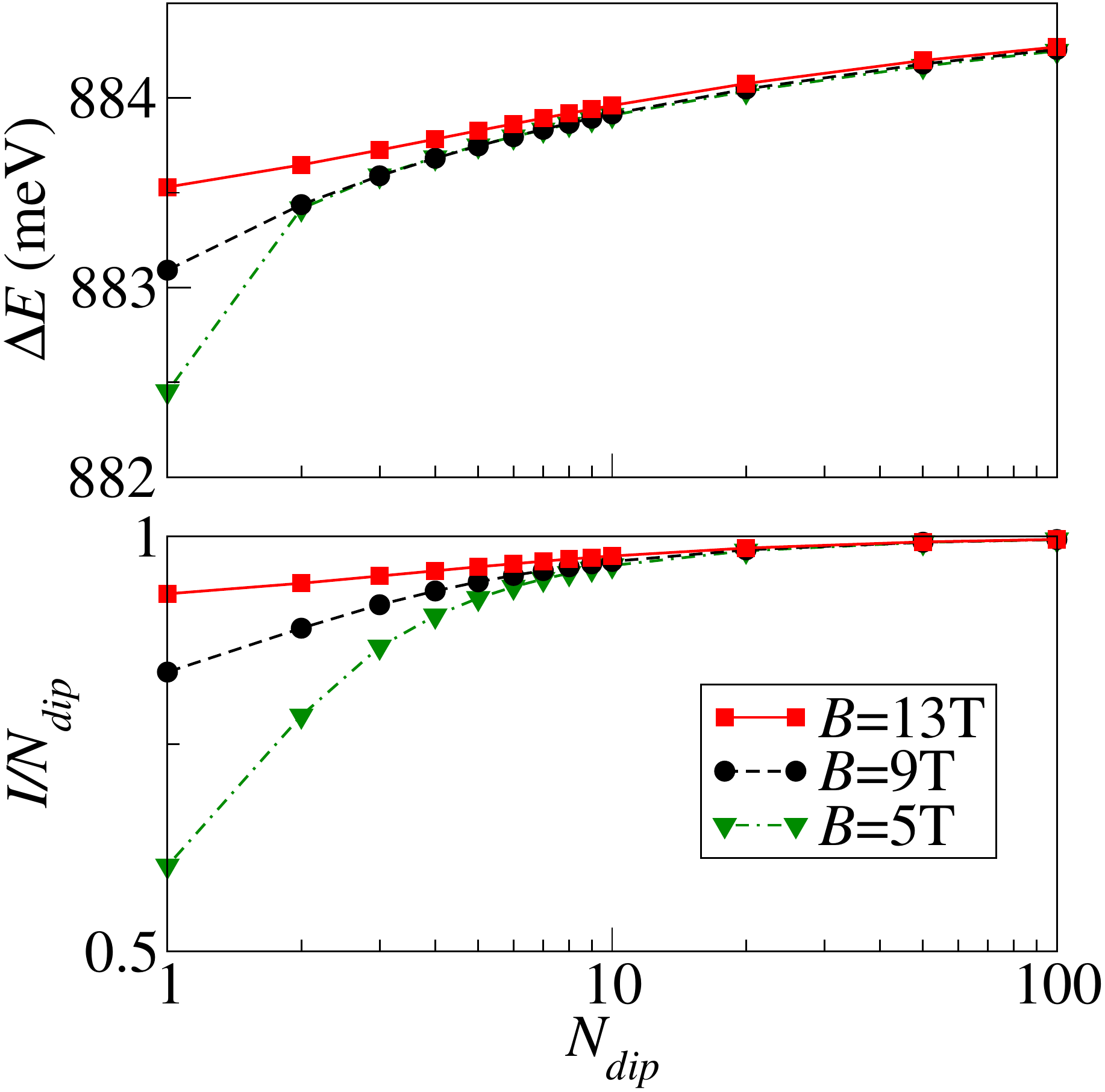}
\caption{(Color Online) Upper panel: Energy position of the emission line as a function of the dipolariton number, $\Delta E=E(N_{dip})-E(N_{dip}-1)$. Lower panel: Relative intensity as a function of the dipolariton number for different magnetic field. $\hbar\omega_h=5$meV, $\hbar\omega_{e1}=7$meV, $\hbar\omega_{e2}=8$meV and $\hbar\omega=884.5$meV.}
\label{fig:emission}
\end{figure}

The position and intensity of the emission line as a function of $N_{dip}$ for several values of $B$ are calculated from Eq. \eqref{eq:emission} and depicted in Fig. \ref{fig:emission}. The number of electron-hole pairs ($N_{exc}$) is restricted to a maximum of 1, this situation is usual, for example, in self-assembled QDs. As expected, at the limit $N_{dip}\gg N_{exc}$, light is emitted at bare photon energies and $I\approx N_{dip}$. When $N_{dip}\sim N_{exc}$ the excitons enhance the trapping of light, decreasing the intensity of the emission, and causing a red-shift. At $B=5$T the system is near resonance, the emission is minimum and highly red-shifted. Different values of the magnetic field can be used to control the emission.

\section{Multiexcitonic QDs \label{sec:multiexcitonic}}

Based on the variational method developed for excitons and polaritons in Refs. \onlinecite{Eastham2001,Boris2000,Vinck2008}, we use an extension of a BCS-like state as a trial function:

\begin{figure}[htbp]
\centering
\includegraphics[width=8cm]{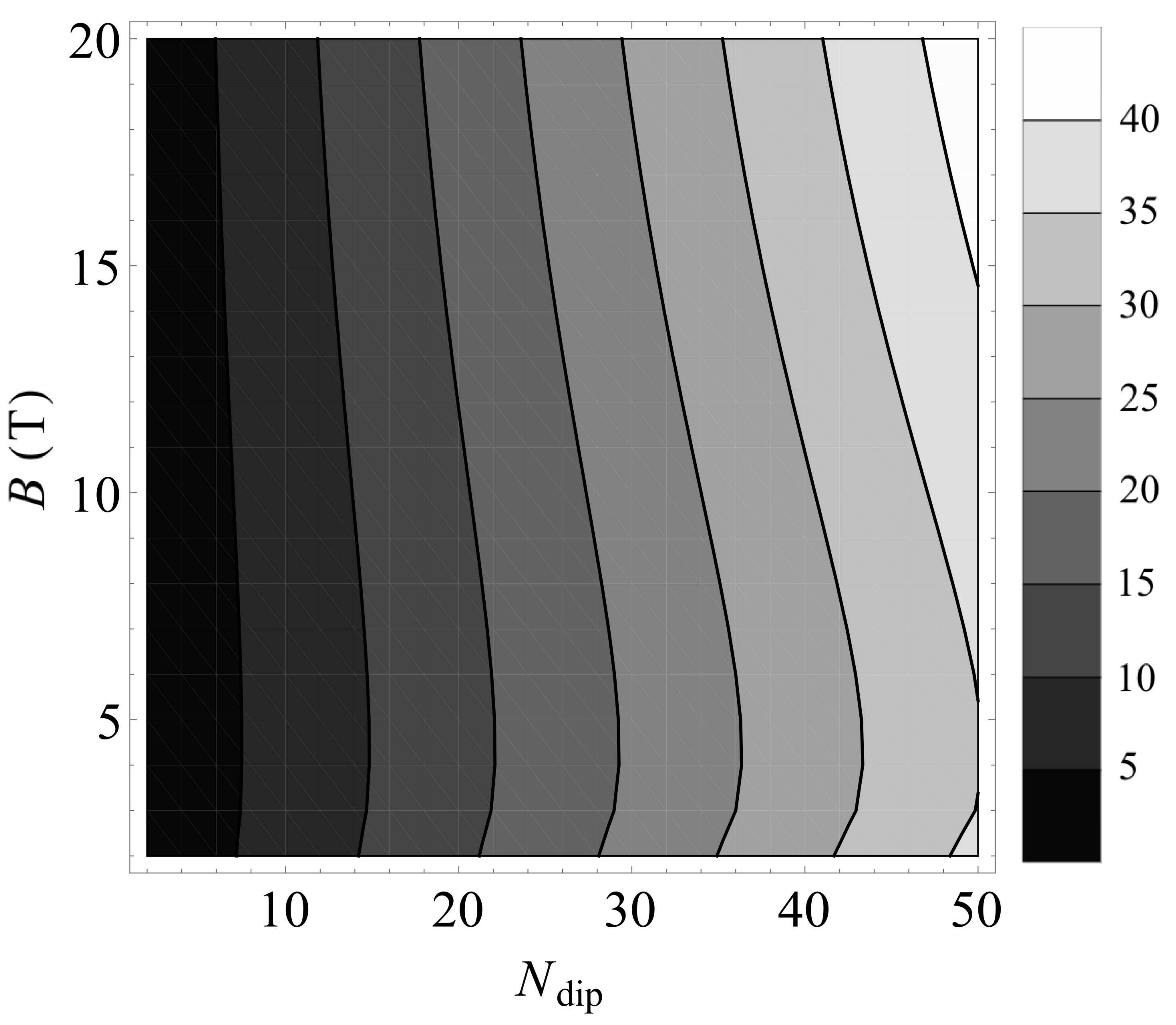}
\caption{(Color Online) Number of photons vs the magnetic field and dipolariton number. Light frequency is set to $\hbar\omega=860$meV and light-matter interaction to $g=1$meV. We consider the QDs to be separated 15nm and assume a tunnelling rate $T_e=2$meV. The offsets are fixed at: $E_0^{e_1}=10$meV and $E_0^h=7$meV, and $E_0^{e_2}$ was taken as in Sec. \ref{sec:exciton}. $\hbar\omega_h=0.5$meV, $\hbar\omega_{e_1}=\hbar\omega_{e_2}=1$meV.}
\label{fig:fotones variacional}
\end{figure}

\begin{equation}
\ket{\sigma,u_n,v_n,w_n}=\ket{\sigma}\otimes\prod_n^{N_{states}}(u_n+v_n\ed{1n}\hd{n}+w_n\ed{2n}\hd{n})\ket{0},
\label{eq:funcion de prueba}
\end{equation}

\noindent where $\ket{\sigma}$ represents a coherent state of light, $u_n^2$ is the probability of not having an electron-hole pair, $v_n^2$ ($w_n^2$) is the probability of having a direct (indirect) pair in the state $n$ and $N_{states}$ is the number of states we consider in our finite basis. The normalization condition of the BCS-like function is $u_n^2+v_n^2+w_n^2=1$.

\begin{figure}[tb]
\centering
\includegraphics[width=9cm]{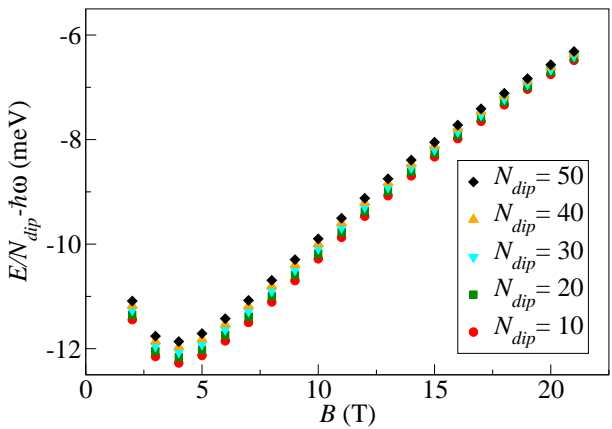}
\caption{(Color Online) Scaling of the total energy as a function of the magnetic field. Parameters are the same as in Fig. \ref{fig:fotones variacional}.}
\label{fig:energia variacional}
\end{figure}

By minimizing the mean value of the hamiltonian $\braket{H}$ calculated using the trial function from Eq. \eqref{eq:funcion de prueba}, with respect to $\sigma$, $v_n$ and $w_n$, imposing a fixed mean number of dipolaritons, we obtain a set of self-consistent equations that are iteratively solved to the desired accuracy. We consider bigger quantum dots that can confine several excitons, this is manifested in smaller values of the parabolic confinements.\cite{Riel2008} 

Fig. \ref{fig:fotones variacional} shows the dependence of the number of photons inside the cavity with the magnetic field and dipolariton number. For a fixed number of dipolaritons, $N_{ph}$ decreases at low values of the magnetic field, when the detuning between matter and light is small. At high values of $B$, the energy of the excitons increase causing the state to be highly photon populated.

The energy of the system as a function of the magnetic field, is displayed in Fig. \ref{fig:energia variacional}. In contrast with Fig. \ref{fig:excitons energies}, for the parameters used the IX has higher energy than the DX, this causes the ground state of the system to have a high component of direct excitons. Therefore, as in Fig. \ref{fig:excitons energies}, the energy decreases at low $B$, but for $B>5$T the total energy of the system rises because the IXs and DXs' energy augments; this change in the slope indicates a switch from paramagnetic to diamagnetic response. As in Ref. \onlinecite{Vinck2008} for polaritons, we find the scaling $E\sim N_{dip}$ as an indicator that the effective interactions between dipolaritons are weak.

We see that the overall behaviour is similar to the usual polariton situation, however, the longer lifetimes and extra control channels give polaritons in QDMs advantages with respect to those in individual QDs.

\section{Conclusions \label{sec:conclusions}}

We have described the properties of a quantum dot molecule embedded in a semiconductor microcavity proposing the formation of dipolaritons in QDs. The latter are found to be similar to their QW analogue but we predict them to exhibit true dark polariton states. The quantum finite system theory developed predicts the magnetic field to be a control parameter for the composition and energy of dipolaritons, and shows that it can be used to control the number of photons inside the cavity and the magnetic response of the system. Our approach is different to what have been done previously with coupled QDs and provides a new perspective for their implementation in quantum devices. Dipolaritons in QDs is an interesting direction for current research in polaritonics given the recent results for their QW analogue ranging from single photon emission to the possible realization of quantum gates, results where the fully quantised nature of QDs may be of great impact.

\acknowledgments

HVP acknowledges the financial support from Colciencias, within the project with code 110156933525, contract number 026-2013, and HERMES code 17432. BAR acknowledges the finantial support from the Universidad de Antioquia through the project CODI-E01620 and the ``Estrategia de Sostenibilidad del Grupo de F\'isica At\'omica y Molecular''.

\bibliography{references.bib}

\begin{thebibliography}{50}%
\makeatletter
\providecommand \@ifxundefined [1]{%
 \@ifx{#1\undefined}
}%
\providecommand \@ifnum [1]{%
 \ifnum #1\expandafter \@firstoftwo
 \else \expandafter \@secondoftwo
 \fi
}%
\providecommand \@ifx [1]{%
 \ifx #1\expandafter \@firstoftwo
 \else \expandafter \@secondoftwo
 \fi
}%
\providecommand \natexlab [1]{#1}%
\providecommand \enquote  [1]{``#1''}%
\providecommand \bibnamefont  [1]{#1}%
\providecommand \bibfnamefont [1]{#1}%
\providecommand \citenamefont [1]{#1}%
\providecommand \href@noop [0]{\@secondoftwo}%
\providecommand \href [0]{\begingroup \@sanitize@url \@href}%
\providecommand \@href[1]{\@@startlink{#1}\@@href}%
\providecommand \@@href[1]{\endgroup#1\@@endlink}%
\providecommand \@sanitize@url [0]{\catcode `\\12\catcode `\$12\catcode
  `\&12\catcode `\#12\catcode `\^12\catcode `\_12\catcode `\%12\relax}%
\providecommand \@@startlink[1]{}%
\providecommand \@@endlink[0]{}%
\providecommand \url  [0]{\begingroup\@sanitize@url \@url }%
\providecommand \@url [1]{\endgroup\@href {#1}{\urlprefix }}%
\providecommand \urlprefix  [0]{URL }%
\providecommand \Eprint [0]{\href }%
\providecommand \doibase [0]{http://dx.doi.org/}%
\providecommand \selectlanguage [0]{\@gobble}%
\providecommand \bibinfo  [0]{\@secondoftwo}%
\providecommand \bibfield  [0]{\@secondoftwo}%
\providecommand \translation [1]{[#1]}%
\providecommand \BibitemOpen [0]{}%
\providecommand \bibitemStop [0]{}%
\providecommand \bibitemNoStop [0]{.\EOS\space}%
\providecommand \EOS [0]{\spacefactor3000\relax}%
\providecommand \BibitemShut  [1]{\csname bibitem#1\endcsname}%
\let\auto@bib@innerbib\@empty
\bibitem [{\citenamefont {Kavokin}\ \emph {et~al.}(2011)\citenamefont
  {Kavokin}, \citenamefont {Baumberg}, \citenamefont {Malpuech},\ and\
  \citenamefont {Laussy}}]{Kavokin}%
  \BibitemOpen
  \bibfield  {author} {\bibinfo {author} {\bibfnamefont {A.}~\bibnamefont
  {Kavokin}}, \bibinfo {author} {\bibfnamefont {J.~J.}\ \bibnamefont
  {Baumberg}}, \bibinfo {author} {\bibfnamefont {G.}~\bibnamefont {Malpuech}},
  \ and\ \bibinfo {author} {\bibfnamefont {F.~P.}\ \bibnamefont {Laussy}},\
  }\href {https://books.google.com.co/books?id=2g7wHcMcaJ0C} {\emph {\bibinfo
  {title} {Microcavities}}},\ Oxford science publications\ (\bibinfo
  {publisher} {OUP Oxford},\ \bibinfo {year} {2011})\BibitemShut {NoStop}%
\bibitem [{\citenamefont {Purcell}(1946)}]{Purcell1946}%
  \BibitemOpen
  \bibfield  {author} {\bibinfo {author} {\bibfnamefont {E.~M.}\ \bibnamefont
  {Purcell}},\ }\href {\doibase 10.1103/PhysRev.69.674.2} {\bibfield  {journal}
  {\bibinfo  {journal} {Phys. Rev.}\ }\textbf {\bibinfo {volume} {69}},\
  \bibinfo {pages} {681} (\bibinfo {year} {1946})}\BibitemShut {NoStop}%
\bibitem [{\citenamefont {Kleppner}(1981)}]{Kleppner1981}%
  \BibitemOpen
  \bibfield  {author} {\bibinfo {author} {\bibfnamefont {D.}~\bibnamefont
  {Kleppner}},\ }\href {\doibase 10.1103/PhysRevLett.47.233} {\bibfield
  {journal} {\bibinfo  {journal} {Phys. Rev. Lett.}\ }\textbf {\bibinfo
  {volume} {47}},\ \bibinfo {pages} {233} (\bibinfo {year} {1981})}\BibitemShut
  {NoStop}%
\bibitem [{\citenamefont {Reithmaier}\ \emph {et~al.}(2004)\citenamefont
  {Reithmaier}, \citenamefont {S{\c{e}}k}, \citenamefont {L{\"o}ffler},
  \citenamefont {Hofmann}, \citenamefont {Kuhn}, \citenamefont {Reitzenstein},
  \citenamefont {Keldysh}, \citenamefont {Kulakovskii}, \citenamefont
  {Reinecke},\ and\ \citenamefont {Forchel}}]{Reithmaier2004}%
  \BibitemOpen
  \bibfield  {author} {\bibinfo {author} {\bibfnamefont {J.}~\bibnamefont
  {Reithmaier}}, \bibinfo {author} {\bibfnamefont {G.}~\bibnamefont
  {S{\c{e}}k}}, \bibinfo {author} {\bibfnamefont {A.}~\bibnamefont
  {L{\"o}ffler}}, \bibinfo {author} {\bibfnamefont {C.}~\bibnamefont
  {Hofmann}}, \bibinfo {author} {\bibfnamefont {S.}~\bibnamefont {Kuhn}},
  \bibinfo {author} {\bibfnamefont {S.}~\bibnamefont {Reitzenstein}}, \bibinfo
  {author} {\bibfnamefont {L.}~\bibnamefont {Keldysh}}, \bibinfo {author}
  {\bibfnamefont {V.}~\bibnamefont {Kulakovskii}}, \bibinfo {author}
  {\bibfnamefont {T.}~\bibnamefont {Reinecke}}, \ and\ \bibinfo {author}
  {\bibfnamefont {A.}~\bibnamefont {Forchel}},\ }\href {\doibase
  10.1038/nature02969} {\bibfield  {journal} {\bibinfo  {journal} {Nature}\
  }\textbf {\bibinfo {volume} {432}},\ \bibinfo {pages} {197} (\bibinfo {year}
  {2004})}\BibitemShut {NoStop}%
\bibitem [{\citenamefont {Robin}\ \emph {et~al.}(2005)\citenamefont {Robin},
  \citenamefont {Andr\'e}, \citenamefont {Balocchi}, \citenamefont {Carayon},
  \citenamefont {Moehl}, \citenamefont {G\'erard},\ and\ \citenamefont
  {Ferlazzo}}]{Robin2005}%
  \BibitemOpen
  \bibfield  {author} {\bibinfo {author} {\bibfnamefont {I.~C.}\ \bibnamefont
  {Robin}}, \bibinfo {author} {\bibfnamefont {R.}~\bibnamefont {Andr\'e}},
  \bibinfo {author} {\bibfnamefont {A.}~\bibnamefont {Balocchi}}, \bibinfo
  {author} {\bibfnamefont {S.}~\bibnamefont {Carayon}}, \bibinfo {author}
  {\bibfnamefont {S.}~\bibnamefont {Moehl}}, \bibinfo {author} {\bibfnamefont
  {J.~M.}\ \bibnamefont {G\'erard}}, \ and\ \bibinfo {author} {\bibfnamefont
  {L.}~\bibnamefont {Ferlazzo}},\ }\href {\doibase 10.1063/1.2136433}
  {\bibfield  {journal} {\bibinfo  {journal} {Applied Physics Letters}\
  }\textbf {\bibinfo {volume} {87}},\ \bibinfo {eid} {233114} (\bibinfo {year}
  {2005})}\BibitemShut {NoStop}%
\bibitem [{\citenamefont {Reitzenstein}\ \emph {et~al.}(2007)\citenamefont
  {Reitzenstein}, \citenamefont {Hofmann}, \citenamefont {Gorbunov},
  \citenamefont {Strau{\ss}}, \citenamefont {Kwon}, \citenamefont {Schneider},
  \citenamefont {L{\"o}ffler}, \citenamefont {H{\"o}fling}, \citenamefont
  {Kamp},\ and\ \citenamefont {Forchel}}]{Reitzenstein2007}%
  \BibitemOpen
  \bibfield  {author} {\bibinfo {author} {\bibfnamefont {S.}~\bibnamefont
  {Reitzenstein}}, \bibinfo {author} {\bibfnamefont {C.}~\bibnamefont
  {Hofmann}}, \bibinfo {author} {\bibfnamefont {A.}~\bibnamefont {Gorbunov}},
  \bibinfo {author} {\bibfnamefont {M.}~\bibnamefont {Strau{\ss}}}, \bibinfo
  {author} {\bibfnamefont {S.~H.}\ \bibnamefont {Kwon}}, \bibinfo {author}
  {\bibfnamefont {C.}~\bibnamefont {Schneider}}, \bibinfo {author}
  {\bibfnamefont {A.}~\bibnamefont {L{\"o}ffler}}, \bibinfo {author}
  {\bibfnamefont {S.}~\bibnamefont {H{\"o}fling}}, \bibinfo {author}
  {\bibfnamefont {M.}~\bibnamefont {Kamp}}, \ and\ \bibinfo {author}
  {\bibfnamefont {A.}~\bibnamefont {Forchel}},\ }\href {\doibase
  10.1063/1.2749862} {\bibfield  {journal} {\bibinfo  {journal} {Applied
  Physics Letters}\ }\textbf {\bibinfo {volume} {90}},\ \bibinfo {eid} {251109}
  (\bibinfo {year} {2007})}\BibitemShut {NoStop}%
\bibitem [{\citenamefont {Englund}\ \emph {et~al.}(2009)\citenamefont
  {Englund}, \citenamefont {Fushman}, \citenamefont {Faraon},\ and\
  \citenamefont {Vu{\v{c}}kovi\'c}}]{Englund2009}%
  \BibitemOpen
  \bibfield  {author} {\bibinfo {author} {\bibfnamefont {D.}~\bibnamefont
  {Englund}}, \bibinfo {author} {\bibfnamefont {I.}~\bibnamefont {Fushman}},
  \bibinfo {author} {\bibfnamefont {A.}~\bibnamefont {Faraon}}, \ and\ \bibinfo
  {author} {\bibfnamefont {J.}~\bibnamefont {Vu{\v{c}}kovi\'c}},\ }\href
  {\doibase 10.1016/j.photonics.2008.11.008} {\bibfield  {journal} {\bibinfo
  {journal} {Photonics and Nanostructures - Fundamentals and Applications}\
  }\textbf {\bibinfo {volume} {7}},\ \bibinfo {pages} {56 } (\bibinfo {year}
  {2009})}\BibitemShut {NoStop}%
\bibitem [{\citenamefont {Sanvitto}\ and\ \citenamefont
  {Timofeev}(2012)}]{Sanvitto}%
  \BibitemOpen
  \bibinfo {editor} {\bibfnamefont {D.}~\bibnamefont {Sanvitto}}\ and\ \bibinfo
  {editor} {\bibfnamefont {V.}~\bibnamefont {Timofeev}},\ eds.,\ \href
  {http://link.springer.com/book/10.1007%2F978-3-642-24186-4} {\emph {\bibinfo
  {title} {Exciton Polaritons in Microcavities: New Frontiers}}},\ Springer
  Series in Solid-State Sciences\ (\bibinfo  {publisher} {Springer Berlin
  Heidelberg},\ \bibinfo {year} {2012})\BibitemShut {NoStop}%
\bibitem [{\citenamefont {Laussy}\ \emph {et~al.}(2009)\citenamefont {Laussy},
  \citenamefont {del Valle},\ and\ \citenamefont {Tejedor}}]{Laussy2009}%
  \BibitemOpen
  \bibfield  {author} {\bibinfo {author} {\bibfnamefont {F.~P.}\ \bibnamefont
  {Laussy}}, \bibinfo {author} {\bibfnamefont {E.}~\bibnamefont {del Valle}}, \
  and\ \bibinfo {author} {\bibfnamefont {C.}~\bibnamefont {Tejedor}},\ }\href
  {\doibase 10.1103/PhysRevB.79.235325} {\bibfield  {journal} {\bibinfo
  {journal} {Phys. Rev. B}\ }\textbf {\bibinfo {volume} {79}},\ \bibinfo
  {pages} {235325} (\bibinfo {year} {2009})}\BibitemShut {NoStop}%
\bibitem [{\citenamefont {Byrnes}\ \emph
  {et~al.}(2014{\natexlab{a}})\citenamefont {Byrnes}, \citenamefont {Kim},\
  and\ \citenamefont {Yamamoto}}]{Byrnes2014-2}%
  \BibitemOpen
  \bibfield  {author} {\bibinfo {author} {\bibfnamefont {T.}~\bibnamefont
  {Byrnes}}, \bibinfo {author} {\bibfnamefont {N.~Y.}\ \bibnamefont {Kim}}, \
  and\ \bibinfo {author} {\bibfnamefont {Y.}~\bibnamefont {Yamamoto}},\ }\href
  {\doibase 10.1038/nphys3143} {\bibfield  {journal} {\bibinfo  {journal}
  {Nature Physics}\ }\textbf {\bibinfo {volume} {10}},\ \bibinfo {pages} {803}
  (\bibinfo {year} {2014}{\natexlab{a}})}\BibitemShut {NoStop}%
\bibitem [{\citenamefont {Koehl}\ and\ \citenamefont
  {Nelson}(2001)}]{Koehl2001}%
  \BibitemOpen
  \bibfield  {author} {\bibinfo {author} {\bibfnamefont {R.~M.}\ \bibnamefont
  {Koehl}}\ and\ \bibinfo {author} {\bibfnamefont {K.~A.}\ \bibnamefont
  {Nelson}},\ }\href {\doibase 10.1063/1.1340579} {\bibfield  {journal}
  {\bibinfo  {journal} {The Journal of Chemical Physics}\ }\textbf {\bibinfo
  {volume} {114}},\ \bibinfo {pages} {1443} (\bibinfo {year}
  {2001})}\BibitemShut {NoStop}%
\bibitem [{\citenamefont {Takahara}\ \emph {et~al.}(2005)\citenamefont
  {Takahara}, \citenamefont {Kusunoki},\ and\ \citenamefont
  {Kobayashi}}]{Takahara2005}%
  \BibitemOpen
  \bibfield  {author} {\bibinfo {author} {\bibfnamefont {J.}~\bibnamefont
  {Takahara}}, \bibinfo {author} {\bibfnamefont {F.}~\bibnamefont {Kusunoki}},
  \ and\ \bibinfo {author} {\bibfnamefont {T.}~\bibnamefont {Kobayashi}},\ }in\
  \href {\doibase 10.1117/12.648767} {\emph {\bibinfo {booktitle} {Proc.
  SPIE}}},\ Vol.\ \bibinfo {volume} {6050}\ (\bibinfo {year} {2005})\ pp.\
  \bibinfo {pages} {60500L--1 -- 60500L--13}\BibitemShut {NoStop}%
\bibitem [{\citenamefont {Feurer}\ \emph {et~al.}(2007)\citenamefont {Feurer},
  \citenamefont {Stoyanov}, \citenamefont {Ward}, \citenamefont {Vaughan},
  \citenamefont {Statz},\ and\ \citenamefont {Nelson}}]{Feurer2007}%
  \BibitemOpen
  \bibfield  {author} {\bibinfo {author} {\bibfnamefont {T.}~\bibnamefont
  {Feurer}}, \bibinfo {author} {\bibfnamefont {N.~S.}\ \bibnamefont
  {Stoyanov}}, \bibinfo {author} {\bibfnamefont {D.~W.}\ \bibnamefont {Ward}},
  \bibinfo {author} {\bibfnamefont {J.~C.}\ \bibnamefont {Vaughan}}, \bibinfo
  {author} {\bibfnamefont {E.~R.}\ \bibnamefont {Statz}}, \ and\ \bibinfo
  {author} {\bibfnamefont {K.~A.}\ \bibnamefont {Nelson}},\ }\href {\doibase
  10.1146/annurev.matsci.37.052506.084327} {\bibfield  {journal} {\bibinfo
  {journal} {Annual Review of Materials Research}\ }\textbf {\bibinfo {volume}
  {37}},\ \bibinfo {pages} {317} (\bibinfo {year} {2007})}\BibitemShut
  {NoStop}%
\bibitem [{\citenamefont {Singh}(2009)}]{Singh2009}%
  \BibitemOpen
  \bibfield  {author} {\bibinfo {author} {\bibfnamefont {M.~R.}\ \bibnamefont
  {Singh}},\ }\href {\doibase 10.1103/PhysRevB.80.195303} {\bibfield  {journal}
  {\bibinfo  {journal} {Phys. Rev. B}\ }\textbf {\bibinfo {volume} {80}},\
  \bibinfo {pages} {195303} (\bibinfo {year} {2009})}\BibitemShut {NoStop}%
\bibitem [{\citenamefont {De~Liberato}(2015)}]{DeLiberato2015}%
  \BibitemOpen
  \bibfield  {author} {\bibinfo {author} {\bibfnamefont {S.}~\bibnamefont
  {De~Liberato}},\ }\href {\doibase 10.1103/PhysRevB.92.125433} {\bibfield
  {journal} {\bibinfo  {journal} {Phys. Rev. B}\ }\textbf {\bibinfo {volume}
  {92}},\ \bibinfo {pages} {125433} (\bibinfo {year} {2015})}\BibitemShut
  {NoStop}%
\bibitem [{\citenamefont {Christmann}\ \emph {et~al.}(2010)\citenamefont
  {Christmann}, \citenamefont {Coulson}, \citenamefont {Baumberg},
  \citenamefont {Pelekanos}, \citenamefont {Hatzopoulos}, \citenamefont
  {Tsintzos},\ and\ \citenamefont {Savvidis}}]{Christmann2010}%
  \BibitemOpen
  \bibfield  {author} {\bibinfo {author} {\bibfnamefont {G.}~\bibnamefont
  {Christmann}}, \bibinfo {author} {\bibfnamefont {C.}~\bibnamefont {Coulson}},
  \bibinfo {author} {\bibfnamefont {J.~J.}\ \bibnamefont {Baumberg}}, \bibinfo
  {author} {\bibfnamefont {N.~T.}\ \bibnamefont {Pelekanos}}, \bibinfo {author}
  {\bibfnamefont {Z.}~\bibnamefont {Hatzopoulos}}, \bibinfo {author}
  {\bibfnamefont {S.~I.}\ \bibnamefont {Tsintzos}}, \ and\ \bibinfo {author}
  {\bibfnamefont {P.~G.}\ \bibnamefont {Savvidis}},\ }\href {\doibase
  10.1103/PhysRevB.82.113308} {\bibfield  {journal} {\bibinfo  {journal} {Phys.
  Rev. B}\ }\textbf {\bibinfo {volume} {82}},\ \bibinfo {pages} {113308}
  (\bibinfo {year} {2010})}\BibitemShut {NoStop}%
\bibitem [{\citenamefont {Christmann}\ \emph {et~al.}(2011)\citenamefont
  {Christmann}, \citenamefont {Askitopoulos}, \citenamefont {Deligeorgis},
  \citenamefont {Hatzopoulos}, \citenamefont {Tsintzos}, \citenamefont
  {Savvidis},\ and\ \citenamefont {Baumberg}}]{Christmann2011}%
  \BibitemOpen
  \bibfield  {author} {\bibinfo {author} {\bibfnamefont {G.}~\bibnamefont
  {Christmann}}, \bibinfo {author} {\bibfnamefont {A.}~\bibnamefont
  {Askitopoulos}}, \bibinfo {author} {\bibfnamefont {G.}~\bibnamefont
  {Deligeorgis}}, \bibinfo {author} {\bibfnamefont {Z.}~\bibnamefont
  {Hatzopoulos}}, \bibinfo {author} {\bibfnamefont {S.~I.}\ \bibnamefont
  {Tsintzos}}, \bibinfo {author} {\bibfnamefont {P.~G.}\ \bibnamefont
  {Savvidis}}, \ and\ \bibinfo {author} {\bibfnamefont {J.~J.}\ \bibnamefont
  {Baumberg}},\ }\href {\doibase 10.1063/1.3559909} {\bibfield  {journal}
  {\bibinfo  {journal} {Applied Physics Letters}\ }\textbf {\bibinfo {volume}
  {98}},\ \bibinfo {eid} {081111} (\bibinfo {year} {2011})}\BibitemShut
  {NoStop}%
\bibitem [{\citenamefont {Cristofolini}\ \emph {et~al.}(2012)\citenamefont
  {Cristofolini}, \citenamefont {Christmann}, \citenamefont {Tsintzos},
  \citenamefont {Deligeorgis}, \citenamefont {Konstantinidis}, \citenamefont
  {Hatzopoulos}, \citenamefont {Savvidis},\ and\ \citenamefont
  {Baumberg}}]{Cristofolini2012}%
  \BibitemOpen
  \bibfield  {author} {\bibinfo {author} {\bibfnamefont {P.}~\bibnamefont
  {Cristofolini}}, \bibinfo {author} {\bibfnamefont {G.}~\bibnamefont
  {Christmann}}, \bibinfo {author} {\bibfnamefont {S.}~\bibnamefont
  {Tsintzos}}, \bibinfo {author} {\bibfnamefont {G.}~\bibnamefont
  {Deligeorgis}}, \bibinfo {author} {\bibfnamefont {G.}~\bibnamefont
  {Konstantinidis}}, \bibinfo {author} {\bibfnamefont {Z.}~\bibnamefont
  {Hatzopoulos}}, \bibinfo {author} {\bibfnamefont {P.}~\bibnamefont
  {Savvidis}}, \ and\ \bibinfo {author} {\bibfnamefont {J.}~\bibnamefont
  {Baumberg}},\ }\href {\doibase 10.1126/science.1219010} {\bibfield  {journal}
  {\bibinfo  {journal} {Science}\ }\textbf {\bibinfo {volume} {336}},\ \bibinfo
  {pages} {704} (\bibinfo {year} {2012})}\BibitemShut {NoStop}%
\bibitem [{\citenamefont {Szyma{\'n}ska}(2012)}]{Szymanska2012}%
  \BibitemOpen
  \bibfield  {author} {\bibinfo {author} {\bibfnamefont {M.~H.}\ \bibnamefont
  {Szyma{\'n}ska}},\ }\href {\doibase 10.1126/science.1221416} {\bibfield
  {journal} {\bibinfo  {journal} {Science}\ }\textbf {\bibinfo {volume}
  {336}},\ \bibinfo {pages} {679} (\bibinfo {year} {2012})}\BibitemShut
  {NoStop}%
\bibitem [{\citenamefont {Kristinsson}\ \emph {et~al.}(2013)\citenamefont
  {Kristinsson}, \citenamefont {Kyriienko}, \citenamefont {Liew},\ and\
  \citenamefont {Shelykh}}]{Kristinsson2013}%
  \BibitemOpen
  \bibfield  {author} {\bibinfo {author} {\bibfnamefont {K.}~\bibnamefont
  {Kristinsson}}, \bibinfo {author} {\bibfnamefont {O.}~\bibnamefont
  {Kyriienko}}, \bibinfo {author} {\bibfnamefont {T.~C.~H.}\ \bibnamefont
  {Liew}}, \ and\ \bibinfo {author} {\bibfnamefont {I.~A.}\ \bibnamefont
  {Shelykh}},\ }\href {\doibase 10.1103/PhysRevB.88.245303} {\bibfield
  {journal} {\bibinfo  {journal} {Phys. Rev. B}\ }\textbf {\bibinfo {volume}
  {88}},\ \bibinfo {pages} {245303} (\bibinfo {year} {2013})}\BibitemShut
  {NoStop}%
\bibitem [{\citenamefont {Kyriienko}\ \emph {et~al.}(2013)\citenamefont
  {Kyriienko}, \citenamefont {Kavokin},\ and\ \citenamefont
  {Shelykh}}]{Kyriienko2013}%
  \BibitemOpen
  \bibfield  {author} {\bibinfo {author} {\bibfnamefont {O.}~\bibnamefont
  {Kyriienko}}, \bibinfo {author} {\bibfnamefont {A.~V.}\ \bibnamefont
  {Kavokin}}, \ and\ \bibinfo {author} {\bibfnamefont {I.~A.}\ \bibnamefont
  {Shelykh}},\ }\href {\doibase 10.1103/PhysRevLett.111.176401} {\bibfield
  {journal} {\bibinfo  {journal} {Phys. Rev. Lett.}\ }\textbf {\bibinfo
  {volume} {111}},\ \bibinfo {pages} {176401} (\bibinfo {year}
  {2013})}\BibitemShut {NoStop}%
\bibitem [{\citenamefont {Kristinsson}\ \emph {et~al.}(2014)\citenamefont
  {Kristinsson}, \citenamefont {Kyriienko},\ and\ \citenamefont
  {Shelykh}}]{Kristinsson2014}%
  \BibitemOpen
  \bibfield  {author} {\bibinfo {author} {\bibfnamefont {K.}~\bibnamefont
  {Kristinsson}}, \bibinfo {author} {\bibfnamefont {O.}~\bibnamefont
  {Kyriienko}}, \ and\ \bibinfo {author} {\bibfnamefont {I.~A.}\ \bibnamefont
  {Shelykh}},\ }\href {\doibase 10.1103/PhysRevA.89.023836} {\bibfield
  {journal} {\bibinfo  {journal} {Phys. Rev. A}\ }\textbf {\bibinfo {volume}
  {89}},\ \bibinfo {pages} {023836} (\bibinfo {year} {2014})}\BibitemShut
  {NoStop}%
\bibitem [{\citenamefont {Kyriienko}\ \emph {et~al.}(2014)\citenamefont
  {Kyriienko}, \citenamefont {Shelykh},\ and\ \citenamefont
  {Liew}}]{Kyriienko2014}%
  \BibitemOpen
  \bibfield  {author} {\bibinfo {author} {\bibfnamefont {O.}~\bibnamefont
  {Kyriienko}}, \bibinfo {author} {\bibfnamefont {I.~A.}\ \bibnamefont
  {Shelykh}}, \ and\ \bibinfo {author} {\bibfnamefont {T.~C.~H.}\ \bibnamefont
  {Liew}},\ }\href {\doibase 10.1103/PhysRevA.90.033807} {\bibfield  {journal}
  {\bibinfo  {journal} {Phys. Rev. A}\ }\textbf {\bibinfo {volume} {90}},\
  \bibinfo {pages} {033807} (\bibinfo {year} {2014})}\BibitemShut {NoStop}%
\bibitem [{\citenamefont {Kyriienko}\ and\ \citenamefont
  {Liew}(2014)}]{Kyriienko2014-2}%
  \BibitemOpen
  \bibfield  {author} {\bibinfo {author} {\bibfnamefont {O.}~\bibnamefont
  {Kyriienko}}\ and\ \bibinfo {author} {\bibfnamefont {T.~C.~H.}\ \bibnamefont
  {Liew}},\ }\href {\doibase 10.1103/PhysRevA.90.063805} {\bibfield  {journal}
  {\bibinfo  {journal} {Phys. Rev. A}\ }\textbf {\bibinfo {volume} {90}},\
  \bibinfo {pages} {063805} (\bibinfo {year} {2014})}\BibitemShut {NoStop}%
\bibitem [{\citenamefont {Byrnes}\ \emph
  {et~al.}(2014{\natexlab{b}})\citenamefont {Byrnes}, \citenamefont {Kolmakov},
  \citenamefont {Kezerashvili},\ and\ \citenamefont {Yamamoto}}]{Byrnes2014}%
  \BibitemOpen
  \bibfield  {author} {\bibinfo {author} {\bibfnamefont {T.}~\bibnamefont
  {Byrnes}}, \bibinfo {author} {\bibfnamefont {G.}~\bibnamefont {Kolmakov}},
  \bibinfo {author} {\bibfnamefont {R.}~\bibnamefont {Kezerashvili}}, \ and\
  \bibinfo {author} {\bibfnamefont {Y.}~\bibnamefont {Yamamoto}},\ }\href
  {\doibase 10.1103/PhysRevB.90.125314} {\bibfield  {journal} {\bibinfo
  {journal} {Phys. Rev. B}\ }\textbf {\bibinfo {volume} {90}},\ \bibinfo
  {pages} {125314} (\bibinfo {year} {2014}{\natexlab{b}})}\BibitemShut
  {NoStop}%
\bibitem [{\citenamefont {Su}\ \emph {et~al.}(2014)\citenamefont {Su},
  \citenamefont {Kim}, \citenamefont {Yamamoto},\ and\ \citenamefont
  {MacDonald}}]{Su2014}%
  \BibitemOpen
  \bibfield  {author} {\bibinfo {author} {\bibfnamefont {J.~J.}\ \bibnamefont
  {Su}}, \bibinfo {author} {\bibfnamefont {N.~Y.}\ \bibnamefont {Kim}},
  \bibinfo {author} {\bibfnamefont {Y.}~\bibnamefont {Yamamoto}}, \ and\
  \bibinfo {author} {\bibfnamefont {A.~H.}\ \bibnamefont {MacDonald}},\ }\href
  {\doibase 10.1103/PhysRevLett.112.116401} {\bibfield  {journal} {\bibinfo
  {journal} {Phys. Rev. Lett.}\ }\textbf {\bibinfo {volume} {112}},\ \bibinfo
  {pages} {116401} (\bibinfo {year} {2014})}\BibitemShut {NoStop}%
\bibitem [{\citenamefont {Coulson}\ \emph {et~al.}(2013)\citenamefont
  {Coulson}, \citenamefont {Christmann}, \citenamefont {Cristofolini},
  \citenamefont {Grossmann}, \citenamefont {Baumberg}, \citenamefont
  {Tsintzos}, \citenamefont {Konstantinidis}, \citenamefont {Hatzopoulos},\
  and\ \citenamefont {Savvidis}}]{Coulson2013}%
  \BibitemOpen
  \bibfield  {author} {\bibinfo {author} {\bibfnamefont {C.}~\bibnamefont
  {Coulson}}, \bibinfo {author} {\bibfnamefont {G.}~\bibnamefont {Christmann}},
  \bibinfo {author} {\bibfnamefont {P.}~\bibnamefont {Cristofolini}}, \bibinfo
  {author} {\bibfnamefont {C.}~\bibnamefont {Grossmann}}, \bibinfo {author}
  {\bibfnamefont {J.~J.}\ \bibnamefont {Baumberg}}, \bibinfo {author}
  {\bibfnamefont {S.~I.}\ \bibnamefont {Tsintzos}}, \bibinfo {author}
  {\bibfnamefont {G.}~\bibnamefont {Konstantinidis}}, \bibinfo {author}
  {\bibfnamefont {Z.}~\bibnamefont {Hatzopoulos}}, \ and\ \bibinfo {author}
  {\bibfnamefont {P.~G.}\ \bibnamefont {Savvidis}},\ }\href {\doibase
  10.1103/PhysRevB.87.045311} {\bibfield  {journal} {\bibinfo  {journal} {Phys.
  Rev. B}\ }\textbf {\bibinfo {volume} {87}},\ \bibinfo {pages} {045311}
  (\bibinfo {year} {2013})}\BibitemShut {NoStop}%
\bibitem [{\citenamefont {Kyriienko}\ and\ \citenamefont
  {Liew}(2016)}]{Kyriienko2016}%
  \BibitemOpen
  \bibfield  {author} {\bibinfo {author} {\bibfnamefont {O.}~\bibnamefont
  {Kyriienko}}\ and\ \bibinfo {author} {\bibfnamefont {T.~C.~H.}\ \bibnamefont
  {Liew}},\ }\href {\doibase 10.1103/PhysRevB.93.035301} {\bibfield  {journal}
  {\bibinfo  {journal} {Phys. Rev. B}\ }\textbf {\bibinfo {volume} {93}},\
  \bibinfo {pages} {035301} (\bibinfo {year} {2016})}\BibitemShut {NoStop}%
\bibitem [{\citenamefont {Khadzhi}\ and\ \citenamefont
  {Vasilieva}(2015)}]{Khadzhi2015}%
  \BibitemOpen
  \bibfield  {author} {\bibinfo {author} {\bibfnamefont {P.}~\bibnamefont
  {Khadzhi}}\ and\ \bibinfo {author} {\bibfnamefont {O.}~\bibnamefont
  {Vasilieva}},\ }\href {\doibase 10.1134/S0021364015210055} {\bibfield
  {journal} {\bibinfo  {journal} {JETP Letters}\ }\textbf {\bibinfo {volume}
  {102}},\ \bibinfo {pages} {581} (\bibinfo {year} {2015})}\BibitemShut
  {NoStop}%
\bibitem [{\citenamefont {Sivalertporn}\ and\ \citenamefont
  {Muljarov}(2015)}]{Sivalertporn2015}%
  \BibitemOpen
  \bibfield  {author} {\bibinfo {author} {\bibfnamefont {K.}~\bibnamefont
  {Sivalertporn}}\ and\ \bibinfo {author} {\bibfnamefont {E.~A.}\ \bibnamefont
  {Muljarov}},\ }\href {\doibase 10.1103/PhysRevLett.115.077401} {\bibfield
  {journal} {\bibinfo  {journal} {Phys. Rev. Lett.}\ }\textbf {\bibinfo
  {volume} {115}},\ \bibinfo {pages} {077401} (\bibinfo {year}
  {2015})}\BibitemShut {NoStop}%
\bibitem [{\citenamefont {Nalitov}\ \emph {et~al.}(2014)\citenamefont
  {Nalitov}, \citenamefont {Solnyshkov}, \citenamefont {Gippius},\ and\
  \citenamefont {Malpuech}}]{Nalitov2014}%
  \BibitemOpen
  \bibfield  {author} {\bibinfo {author} {\bibfnamefont {A.~V.}\ \bibnamefont
  {Nalitov}}, \bibinfo {author} {\bibfnamefont {D.~D.}\ \bibnamefont
  {Solnyshkov}}, \bibinfo {author} {\bibfnamefont {N.~A.}\ \bibnamefont
  {Gippius}}, \ and\ \bibinfo {author} {\bibfnamefont {G.}~\bibnamefont
  {Malpuech}},\ }\href {\doibase 10.1103/PhysRevB.90.235304} {\bibfield
  {journal} {\bibinfo  {journal} {Phys. Rev. B}\ }\textbf {\bibinfo {volume}
  {90}},\ \bibinfo {pages} {235304} (\bibinfo {year} {2014})}\BibitemShut
  {NoStop}%
\bibitem [{\citenamefont {Oosterkamp}\ \emph {et~al.}(1998)\citenamefont
  {Oosterkamp}, \citenamefont {Fujisawa}, \citenamefont {Van Der~Wiel},
  \citenamefont {Ishibashi}, \citenamefont {Hijman}, \citenamefont {Tarucha},\
  and\ \citenamefont {Kouwenhoven}}]{Oosterkamp1998}%
  \BibitemOpen
  \bibfield  {author} {\bibinfo {author} {\bibfnamefont {T.}~\bibnamefont
  {Oosterkamp}}, \bibinfo {author} {\bibfnamefont {T.}~\bibnamefont
  {Fujisawa}}, \bibinfo {author} {\bibfnamefont {W.}~\bibnamefont {Van
  Der~Wiel}}, \bibinfo {author} {\bibfnamefont {K.}~\bibnamefont {Ishibashi}},
  \bibinfo {author} {\bibfnamefont {R.}~\bibnamefont {Hijman}}, \bibinfo
  {author} {\bibfnamefont {S.}~\bibnamefont {Tarucha}}, \ and\ \bibinfo
  {author} {\bibfnamefont {L.}~\bibnamefont {Kouwenhoven}},\ }\href {\doibase
  10.1038/27617} {\bibfield  {journal} {\bibinfo  {journal} {Nature}\ }\textbf
  {\bibinfo {volume} {395}},\ \bibinfo {pages} {873} (\bibinfo {year}
  {1998})}\BibitemShut {NoStop}%
\bibitem [{\citenamefont {Austing}\ \emph {et~al.}(1998)\citenamefont
  {Austing}, \citenamefont {Honda}, \citenamefont {Muraki}, \citenamefont
  {Tokura},\ and\ \citenamefont {Tarucha}}]{Austing1998}%
  \BibitemOpen
  \bibfield  {author} {\bibinfo {author} {\bibfnamefont {D.}~\bibnamefont
  {Austing}}, \bibinfo {author} {\bibfnamefont {T.}~\bibnamefont {Honda}},
  \bibinfo {author} {\bibfnamefont {K.}~\bibnamefont {Muraki}}, \bibinfo
  {author} {\bibfnamefont {Y.}~\bibnamefont {Tokura}}, \ and\ \bibinfo {author}
  {\bibfnamefont {S.}~\bibnamefont {Tarucha}},\ }\href {\doibase
  10.1016/S0921-4526(98)00099-4} {\bibfield  {journal} {\bibinfo  {journal}
  {Physica B: Condensed Matter}\ }\textbf {\bibinfo {volume} {249 - 251}},\
  \bibinfo {pages} {206 } (\bibinfo {year} {1998})}\BibitemShut {NoStop}%
\bibitem [{\citenamefont {Wu}\ and\ \citenamefont {Wang}(2014)}]{Wu}%
  \BibitemOpen
  \bibinfo {editor} {\bibfnamefont {J.}~\bibnamefont {Wu}}\ and\ \bibinfo
  {editor} {\bibfnamefont {Z.~M.}\ \bibnamefont {Wang}},\ eds.,\ \href
  {http://www.springer.com/gb/book/9781461481294} {\emph {\bibinfo {title}
  {Quantum Dot Molecules}}},\ \bibinfo {series} {Lecture Notes in Nanoscale
  Science and Technology}, Vol.~\bibinfo {volume} {14}\ (\bibinfo  {publisher}
  {Springer-Verlag New York},\ \bibinfo {year} {2014})\BibitemShut {NoStop}%
\bibitem [{\citenamefont {Villas-B\^oas}\ \emph
  {et~al.}(2004{\natexlab{a}})\citenamefont {Villas-B\^oas}, \citenamefont
  {Govorov},\ and\ \citenamefont {Ulloa}}]{Villas-Boas2004}%
  \BibitemOpen
  \bibfield  {author} {\bibinfo {author} {\bibfnamefont {J.~M.}\ \bibnamefont
  {Villas-B\^oas}}, \bibinfo {author} {\bibfnamefont {A.~O.}\ \bibnamefont
  {Govorov}}, \ and\ \bibinfo {author} {\bibfnamefont {S.~E.}\ \bibnamefont
  {Ulloa}},\ }\href {\doibase 10.1103/PhysRevB.69.125342} {\bibfield  {journal}
  {\bibinfo  {journal} {Phys. Rev. B}\ }\textbf {\bibinfo {volume} {69}},\
  \bibinfo {pages} {125342} (\bibinfo {year} {2004}{\natexlab{a}})}\BibitemShut
  {NoStop}%
\bibitem [{\citenamefont {Villas-B\^oas}\ \emph
  {et~al.}(2004{\natexlab{b}})\citenamefont {Villas-B\^oas}, \citenamefont
  {Ulloa},\ and\ \citenamefont {Studart}}]{Villas-Boas2004-2}%
  \BibitemOpen
  \bibfield  {author} {\bibinfo {author} {\bibfnamefont {J.~M.}\ \bibnamefont
  {Villas-B\^oas}}, \bibinfo {author} {\bibfnamefont {S.~E.}\ \bibnamefont
  {Ulloa}}, \ and\ \bibinfo {author} {\bibfnamefont {N.}~\bibnamefont
  {Studart}},\ }\href {\doibase 10.1103/PhysRevB.70.041302} {\bibfield
  {journal} {\bibinfo  {journal} {Phys. Rev. B}\ }\textbf {\bibinfo {volume}
  {70}},\ \bibinfo {pages} {041302} (\bibinfo {year}
  {2004}{\natexlab{b}})}\BibitemShut {NoStop}%
\bibitem [{\citenamefont {M\"uller}\ \emph {et~al.}(2012)\citenamefont
  {M\"uller}, \citenamefont {Bechtold}, \citenamefont {Ruppert}, \citenamefont
  {Zecherle}, \citenamefont {Reithmaier}, \citenamefont {Bichler},
  \citenamefont {Krenner}, \citenamefont {Abstreiter}, \citenamefont
  {Holleitner}, \citenamefont {Villas-Boas}, \citenamefont {Betz},\ and\
  \citenamefont {Finley}}]{Muller2012}%
  \BibitemOpen
  \bibfield  {author} {\bibinfo {author} {\bibfnamefont {K.}~\bibnamefont
  {M\"uller}}, \bibinfo {author} {\bibfnamefont {A.}~\bibnamefont {Bechtold}},
  \bibinfo {author} {\bibfnamefont {C.}~\bibnamefont {Ruppert}}, \bibinfo
  {author} {\bibfnamefont {M.}~\bibnamefont {Zecherle}}, \bibinfo {author}
  {\bibfnamefont {G.}~\bibnamefont {Reithmaier}}, \bibinfo {author}
  {\bibfnamefont {M.}~\bibnamefont {Bichler}}, \bibinfo {author} {\bibfnamefont
  {H.}~\bibnamefont {Krenner}}, \bibinfo {author} {\bibfnamefont
  {G.}~\bibnamefont {Abstreiter}}, \bibinfo {author} {\bibfnamefont
  {A.}~\bibnamefont {Holleitner}}, \bibinfo {author} {\bibfnamefont
  {J.}~\bibnamefont {Villas-Boas}}, \bibinfo {author} {\bibfnamefont
  {M.}~\bibnamefont {Betz}}, \ and\ \bibinfo {author} {\bibfnamefont
  {J.}~\bibnamefont {Finley}},\ }\href {\doibase
  10.1103/PhysRevLett.108.197402} {\bibfield  {journal} {\bibinfo  {journal}
  {Phys. Rev. Lett.}\ }\textbf {\bibinfo {volume} {108}},\ \bibinfo {pages}
  {197402} (\bibinfo {year} {2012})}\BibitemShut {NoStop}%
\bibitem [{\citenamefont {M\"uller}\ \emph {et~al.}(2013)\citenamefont
  {M\"uller}, \citenamefont {Bechtold}, \citenamefont {Ruppert}, \citenamefont
  {Kaldewey}, \citenamefont {Zecherle}, \citenamefont {Wildmann}, \citenamefont
  {Bichler}, \citenamefont {Krenner}, \citenamefont {Villas-B\^oas},
  \citenamefont {Abstreiter}, \citenamefont {Betz},\ and\ \citenamefont
  {Finley}}]{Muller2013}%
  \BibitemOpen
  \bibfield  {author} {\bibinfo {author} {\bibfnamefont {K.}~\bibnamefont
  {M\"uller}}, \bibinfo {author} {\bibfnamefont {A.}~\bibnamefont {Bechtold}},
  \bibinfo {author} {\bibfnamefont {C.}~\bibnamefont {Ruppert}}, \bibinfo
  {author} {\bibfnamefont {T.}~\bibnamefont {Kaldewey}}, \bibinfo {author}
  {\bibfnamefont {M.}~\bibnamefont {Zecherle}}, \bibinfo {author}
  {\bibfnamefont {J.~S.}\ \bibnamefont {Wildmann}}, \bibinfo {author}
  {\bibfnamefont {M.}~\bibnamefont {Bichler}}, \bibinfo {author} {\bibfnamefont
  {H.~J.}\ \bibnamefont {Krenner}}, \bibinfo {author} {\bibfnamefont {J.~M.}\
  \bibnamefont {Villas-B\^oas}}, \bibinfo {author} {\bibfnamefont
  {G.}~\bibnamefont {Abstreiter}}, \bibinfo {author} {\bibfnamefont
  {M.}~\bibnamefont {Betz}}, \ and\ \bibinfo {author} {\bibfnamefont {J.~J.}\
  \bibnamefont {Finley}},\ }\href {\doibase 10.1002/andp.201200195} {\bibfield
  {journal} {\bibinfo  {journal} {Annalen der Physik}\ }\textbf {\bibinfo
  {volume} {525}},\ \bibinfo {pages} {49} (\bibinfo {year} {2013})}\BibitemShut
  {NoStop}%
\bibitem [{\citenamefont {Rojas-Arias}\ \emph {et~al.}(2015)\citenamefont
  {Rojas-Arias}, \citenamefont {{Brice\~no}-Villalba},\ and\ \citenamefont
  {Vinck-Posada}}]{Rojas-Arias2015}%
  \BibitemOpen
  \bibfield  {author} {\bibinfo {author} {\bibfnamefont {J.~S.}\ \bibnamefont
  {Rojas-Arias}}, \bibinfo {author} {\bibfnamefont {L.~A.}\ \bibnamefont
  {{Brice\~no}-Villalba}}, \ and\ \bibinfo {author} {\bibfnamefont
  {H.}~\bibnamefont {Vinck-Posada}},\ }in\ \href {\doibase
  10.1364/IPRSN.2015.JM3A.20} {\emph {\bibinfo {booktitle} {Advanced Photonics
  2015}}}\ (\bibinfo  {publisher} {Optical Society of America},\ \bibinfo
  {year} {2015})\ p.\ \bibinfo {pages} {JM3A.20}\BibitemShut {NoStop}%
\bibitem [{\citenamefont {Vinck}\ \emph {et~al.}(2006)\citenamefont {Vinck},
  \citenamefont {Rodriguez},\ and\ \citenamefont {Gonzalez}}]{Vinck2006}%
  \BibitemOpen
  \bibfield  {author} {\bibinfo {author} {\bibfnamefont {H.}~\bibnamefont
  {Vinck}}, \bibinfo {author} {\bibfnamefont {B.}~\bibnamefont {Rodriguez}}, \
  and\ \bibinfo {author} {\bibfnamefont {A.}~\bibnamefont {Gonzalez}},\ }\href
  {\doibase 10.1016/j.physe.2006.06.019} {\bibfield  {journal} {\bibinfo
  {journal} {Physica E: Low-dimensional Systems and Nanostructures}\ }\textbf
  {\bibinfo {volume} {35}},\ \bibinfo {pages} {99 } (\bibinfo {year}
  {2006})}\BibitemShut {NoStop}%
\bibitem [{\citenamefont {Riel}(2008)}]{Riel2008}%
  \BibitemOpen
  \bibfield  {author} {\bibinfo {author} {\bibfnamefont {B.}~\bibnamefont
  {Riel}},\ }\href {\doibase 10.1119/1.2907856} {\bibfield  {journal} {\bibinfo
   {journal} {American Journal of Physics}\ }\textbf {\bibinfo {volume} {76}},\
  \bibinfo {pages} {750} (\bibinfo {year} {2008})}\BibitemShut {NoStop}%
\bibitem [{\citenamefont {Mondragon-Shem}\ \emph {et~al.}(2010)\citenamefont
  {Mondragon-Shem}, \citenamefont {Rodr\'iguez},\ and\ \citenamefont
  {L\'opez}}]{Mondragon-Shem2010}%
  \BibitemOpen
  \bibfield  {author} {\bibinfo {author} {\bibfnamefont {I.}~\bibnamefont
  {Mondragon-Shem}}, \bibinfo {author} {\bibfnamefont {B.~A.}\ \bibnamefont
  {Rodr\'iguez}}, \ and\ \bibinfo {author} {\bibfnamefont {F.~E.}\ \bibnamefont
  {L\'opez}},\ }\href {\doibase 10.1016/j.cpc.2010.05.001} {\bibfield
  {journal} {\bibinfo  {journal} {Computer Physics Communications}\ }\textbf
  {\bibinfo {volume} {181}},\ \bibinfo {pages} {1510 } (\bibinfo {year}
  {2010})}\BibitemShut {NoStop}%
\bibitem [{\citenamefont {Smoliner}\ \emph {et~al.}(1991)\citenamefont
  {Smoliner}, \citenamefont {Demmerle}, \citenamefont {Hirler}, \citenamefont
  {Gornik}, \citenamefont {Weimann}, \citenamefont {Hauser},\ and\
  \citenamefont {Schlapp}}]{Smoliner1991}%
  \BibitemOpen
  \bibfield  {author} {\bibinfo {author} {\bibfnamefont {J.}~\bibnamefont
  {Smoliner}}, \bibinfo {author} {\bibfnamefont {W.}~\bibnamefont {Demmerle}},
  \bibinfo {author} {\bibfnamefont {F.}~\bibnamefont {Hirler}}, \bibinfo
  {author} {\bibfnamefont {E.}~\bibnamefont {Gornik}}, \bibinfo {author}
  {\bibfnamefont {G.}~\bibnamefont {Weimann}}, \bibinfo {author} {\bibfnamefont
  {M.}~\bibnamefont {Hauser}}, \ and\ \bibinfo {author} {\bibfnamefont
  {W.}~\bibnamefont {Schlapp}},\ }\href {\doibase 10.1103/PhysRevB.43.7358}
  {\bibfield  {journal} {\bibinfo  {journal} {Phys. Rev. B}\ }\textbf {\bibinfo
  {volume} {43}},\ \bibinfo {pages} {7358} (\bibinfo {year}
  {1991})}\BibitemShut {NoStop}%
\bibitem [{\citenamefont {Tarucha}\ \emph {et~al.}(1999)\citenamefont
  {Tarucha}, \citenamefont {Fujisawa}, \citenamefont {Ono}, \citenamefont
  {Austing}, \citenamefont {Oosterkamp}, \citenamefont {van~der Wiel},\ and\
  \citenamefont {Kouwenhoven}}]{Tarucha1999}%
  \BibitemOpen
  \bibfield  {author} {\bibinfo {author} {\bibfnamefont {S.}~\bibnamefont
  {Tarucha}}, \bibinfo {author} {\bibfnamefont {T.}~\bibnamefont {Fujisawa}},
  \bibinfo {author} {\bibfnamefont {K.}~\bibnamefont {Ono}}, \bibinfo {author}
  {\bibfnamefont {D.}~\bibnamefont {Austing}}, \bibinfo {author} {\bibfnamefont
  {T.}~\bibnamefont {Oosterkamp}}, \bibinfo {author} {\bibfnamefont
  {W.}~\bibnamefont {van~der Wiel}}, \ and\ \bibinfo {author} {\bibfnamefont
  {L.}~\bibnamefont {Kouwenhoven}},\ }\href {\doibase
  10.1016/S0167-9317(99)00162-8} {\bibfield  {journal} {\bibinfo  {journal}
  {Microelectronic Engineering}\ }\textbf {\bibinfo {volume} {47}},\ \bibinfo
  {pages} {101 } (\bibinfo {year} {1999})}\BibitemShut {NoStop}%
\bibitem [{\citenamefont {Vera}\ \emph {et~al.}(2009)\citenamefont {Vera},
  \citenamefont {Vinck-Posada},\ and\ \citenamefont {Gonz\'alez}}]{Vera2009}%
  \BibitemOpen
  \bibfield  {author} {\bibinfo {author} {\bibfnamefont {C.}~\bibnamefont
  {Vera}}, \bibinfo {author} {\bibfnamefont {H.}~\bibnamefont {Vinck-Posada}},
  \ and\ \bibinfo {author} {\bibfnamefont {A.}~\bibnamefont {Gonz\'alez}},\
  }\href {\doibase 10.1103/PhysRevB.80.125302} {\bibfield  {journal} {\bibinfo
  {journal} {Phys. Rev. B}\ }\textbf {\bibinfo {volume} {80}},\ \bibinfo
  {pages} {125302} (\bibinfo {year} {2009})}\BibitemShut {NoStop}%
\bibitem [{\citenamefont {Eastham}\ and\ \citenamefont
  {Littlewood}(2001)}]{Eastham2001}%
  \BibitemOpen
  \bibfield  {author} {\bibinfo {author} {\bibfnamefont {P.}~\bibnamefont
  {Eastham}}\ and\ \bibinfo {author} {\bibfnamefont {P.}~\bibnamefont
  {Littlewood}},\ }\href {\doibase 10.1103/PhysRevB.64.235101} {\bibfield
  {journal} {\bibinfo  {journal} {Phys. Rev. B}\ }\textbf {\bibinfo {volume}
  {64}},\ \bibinfo {pages} {235101} (\bibinfo {year} {2001})}\BibitemShut
  {NoStop}%
\bibitem [{\citenamefont {Vinck-Posada}\ \emph {et~al.}(2007)\citenamefont
  {Vinck-Posada}, \citenamefont {Rodriguez}, \citenamefont {Guimaraes},
  \citenamefont {Cabo},\ and\ \citenamefont {Gonzalez}}]{Vinck2007}%
  \BibitemOpen
  \bibfield  {author} {\bibinfo {author} {\bibfnamefont {H.}~\bibnamefont
  {Vinck-Posada}}, \bibinfo {author} {\bibfnamefont {B.}~\bibnamefont
  {Rodriguez}}, \bibinfo {author} {\bibfnamefont {P.}~\bibnamefont
  {Guimaraes}}, \bibinfo {author} {\bibfnamefont {A.}~\bibnamefont {Cabo}}, \
  and\ \bibinfo {author} {\bibfnamefont {A.}~\bibnamefont {Gonzalez}},\ }\href
  {\doibase 10.1103/PhysRevLett.98.167405} {\bibfield  {journal} {\bibinfo
  {journal} {Phys. Rev. Lett.}\ }\textbf {\bibinfo {volume} {98}},\ \bibinfo
  {pages} {167405} (\bibinfo {year} {2007})}\BibitemShut {NoStop}%
\bibitem [{\citenamefont {Vinck-Posada}\ \emph {et~al.}(2008)\citenamefont
  {Vinck-Posada}, \citenamefont {Rodriguez},\ and\ \citenamefont
  {Gonzalez}}]{Vinck2008}%
  \BibitemOpen
  \bibfield  {author} {\bibinfo {author} {\bibfnamefont {H.}~\bibnamefont
  {Vinck-Posada}}, \bibinfo {author} {\bibfnamefont {B.}~\bibnamefont
  {Rodriguez}}, \ and\ \bibinfo {author} {\bibfnamefont {A.}~\bibnamefont
  {Gonzalez}},\ }\href {\doibase 10.1016/j.spmi.2007.07.023} {\bibfield
  {journal} {\bibinfo  {journal} {Superlattices and Microstructures}\ }\textbf
  {\bibinfo {volume} {43}},\ \bibinfo {pages} {500 } (\bibinfo {year}
  {2008})}\BibitemShut {NoStop}%
\bibitem [{\citenamefont {{Su{\'a}rez-Forero}}\ \emph
  {et~al.}(2012)\citenamefont {{Su{\'a}rez-Forero}}, \citenamefont
  {{Cipagauta}}, \citenamefont {{Vinck-Posada}}, \citenamefont
  {{Fonseca-Romero}},\ and\ \citenamefont {{Rodr{\'{\i}}guez}}}]{Suarez2012}%
  \BibitemOpen
  \bibfield  {author} {\bibinfo {author} {\bibfnamefont {D.~G.}\ \bibnamefont
  {{Su{\'a}rez-Forero}}}, \bibinfo {author} {\bibfnamefont {G.}~\bibnamefont
  {{Cipagauta}}}, \bibinfo {author} {\bibfnamefont {H.}~\bibnamefont
  {{Vinck-Posada}}}, \bibinfo {author} {\bibfnamefont {K.~M.}\ \bibnamefont
  {{Fonseca-Romero}}}, \ and\ \bibinfo {author} {\bibfnamefont {B.~A.}\
  \bibnamefont {{Rodr{\'{\i}}guez}}},\ }\href@noop {} {\  (\bibinfo {year}
  {2012})},\ \Eprint {http://arxiv.org/abs/1205.2719} {arXiv:1205.2719}
  \BibitemShut {NoStop}%
\bibitem [{\citenamefont {Rodr\'iguez}\ \emph {et~al.}(2000)\citenamefont
  {Rodr\'iguez}, \citenamefont {Gonz\'alez}, \citenamefont {Quiroga},
  \citenamefont {Rodr\'iguez},\ and\ \citenamefont {Capote}}]{Boris2000}%
  \BibitemOpen
  \bibfield  {author} {\bibinfo {author} {\bibfnamefont {B.}~\bibnamefont
  {Rodr\'iguez}}, \bibinfo {author} {\bibfnamefont {A.}~\bibnamefont
  {Gonz\'alez}}, \bibinfo {author} {\bibfnamefont {L.}~\bibnamefont {Quiroga}},
  \bibinfo {author} {\bibfnamefont {F.}~\bibnamefont {Rodr\'iguez}}, \ and\
  \bibinfo {author} {\bibfnamefont {R.}~\bibnamefont {Capote}},\ }\href
  {\doibase 10.1142/S0217979200000078} {\bibfield  {journal} {\bibinfo
  {journal} {International Journal of Modern Physics B}\ }\textbf {\bibinfo
  {volume} {14}},\ \bibinfo {pages} {71} (\bibinfo {year} {2000})}\BibitemShut
  {NoStop}%
\end{thebibliography}%

\end{document}